\documentclass[traditabstract]{aa}
\usepackage{graphicx}
\usepackage{txfonts}
\usepackage{epsfig}
\usepackage{natbib}

   % ...
   % ...
 % ...

 \usepackage{psboxit}

                {\bfseries}%
                {}

\usepackage{psboxit}
 \PScommands
 \newenvironment{Note}%
                {\noindent\normalfont\normalsize%\itshape%
                 \begin{boxitpara}{box 0.85 setgray fill}%
                }%
                {\end{boxitpara}}

\newenvironment{erase}%
                {\noindent\normalfont\normalsize%\itshape%
                 \begin{boxitpara}{box 0.35 setgray fill}%
                }%
                {\end{boxitpara}}

\newcommand{\bnote}{\begin{Note}}
\newcommand{\enote}{\end{Note}}

\newcommand{\berase}{\begin{erase}}
\newcommand{\eerase}{\end{erase}}

 \usepackage{color}
 \usepackage{pstcol}

\begin{document}

\title{Benchmarking atomic data for astrophysics: \\
a first look at the  soft X-ray lines}

\author{G. Del Zanna} 
\institute{DAMTP, Centre for Mathematical Sciences,  Wilberforce Road, Cambridge, CB3 0WA, UK}

%\offprints{G. Del Zanna  (gdz@mssl.ucl.ac.uk)}
  \date{Received  29 June 2012  ; accepted 7 August 2012 } %%%%% 30 September 2011

  \abstract{A collection of the best solar and laboratory spectra in the soft X-rays is used here
to perform a preliminary  benchmark in this wavelength region,
by comparing observed vs. predicted wavelengths and calibrated solar irradiances.
The benchmark focuses on the \ion{Fe}{ix} --  \ion{Fe}{xiv} ions,
for which we have recently calculated the relevant atomic data, however a few other ions
 have also been benchmarked.
The iron ions are dominating the soft X-rays, however 
a large fraction of the strongest
soft X-ray lines due to $n=4 \to n=3$ transitions were previously unidentified. 
The strongest transitions are all identified here, in particular 
the decays from the core-excited levels (3s 3p$^l$ 4s, $l=$ 5,4,3,2,1 for 
\ion{Fe}{x}, \ion{Fe}{xi}, \ion{Fe}{xii}, \ion{Fe}{xiii}, and  \ion{Fe}{xiv}
respectively) which are the strongest soft X-ray transitions from these ions.
Many new identifications are proposed, some only tentatively.
Good agreement in terms of solar irradiances between the soft-Xray and EUV
($n=3 \to n=3$) transitions is found, confirming the reliability
of the new large-scale calculations. 
Some of the new atomic data and identifications are particularly important
for the  Solar Dynamic  Observatory (SDO)  Atmospheric Imaging Assembly 
(AIA) 94~\AA\ band. 
}

\titlerunning{Benchmarking soft X-ray lines}

\maketitle

\section{Introduction }

The soft X-ray (50--170~\AA) spectrum
 is rich in $n=4 \to n=3$ transitions from  highly ionised iron ions,
from \ion{Fe}{viii} to \ion{Fe}{xvi} 
(see, e.g.  \citealt{fawcett_etal:72}, \citealt{manson:72}, and
\citealt{behring_etal:72}). 
Various current missions are routinely observing the  soft X-rays.
For example, Chandra with the LETG, and the  Solar Dynamic  Observatory
(SDO) with a suite of instruments.
The SDO  Extreme ultraviolet Variability Experiment (EVE)
\citep{woods_etal:12} has  been providing soft  X-ray irradiances long-ward of 60~\AA, 
while the Atmospheric Imaging Assembly 
(AIA, see \citealt{lemen_etal:12}) has been  
observing, for the first time routinely,  the 
solar corona in two broad-bands centred in the soft X-rays, around  94 and 131~\AA.

Very little atomic data were available in the soft X-rays
and the majority of the spectral lines still await firm identification.
Within the APAP network (www.apap-network.org), we are carrying out a long-term
project for calculating accurate atomic data for the soft X-rays.
We started with the  \ion{Fe}{viii}--\ion{Fe}{xiv} iron ions.
The atomic data for \ion{Fe}{viii} and \ion{Fe}{ix} have recently been discussed in
\cite{odwyer_etal:11_fe_9}, where new DW calculations for these two ions were presented.   
The main problems  related to calculating accurate atomic data  for the $n=4$ levels
are discussed in \cite{delzanna_etal:12_fe_10}, where new large-scale 
R-matrix atomic calculations 
for \ion{Fe}{x} have been presented.
A similar work on \ion{Fe}{xi}, \ion{Fe}{xii}, and \ion{Fe}{xiii}
has been presented in \cite{delzanna_storey:12_fe_11, delzanna_etal:12_fe_12,delzanna_storey:12_fe_13}.
New  atomic data for \ion{Fe}{xiv} and \ion{Fe}{xvi} have also recently 
been calculated with the R-matrix  method
 by \cite{liang_etal:10_fe_14} and \cite{liang_etal:09_na-like}.

It is therefore now possible to provide the first benchmark study 
for the soft X-rays for these iron ions, based on accurate atomic calculations.
Previously,  \cite{lepson_etal:02} provided some tentative identifications for 
\ion{Fe}{vii} -- \ion{Fe}{x} based on EBIT laboratory measurements and 
unpublished  distorted wave (DW)  calculations.
\cite{liang_zhao:10} discussed \ion{Fe}{ix} -- \ion{Fe}{xvi}
emission lines using DW  calculations obtained with the Flexible Atomic Code (FAC)
and Chandra LETG observations of Procyon. However, various problems with 
this work have been found. First,  almost all of their identifications were either
previously known or are at odds with the present results. 
Second, large discrepancies
between observed and predicted line fluxes were present. Third, the 
Procyon spectra were poor in terms of signal and spectral resolution,
when compared to the solar spectra used in the present benchmark.

Recently, \cite{testa_etal:12} also used Chandra LETG observations of Procyon
to benchmark CHIANTI v.6 \citep{dere_etal:97,dere_etal:09_chianti_v6} data, 
however no atomic data for the 
 \ion{Fe}{x} -- \ion{Fe}{xiv} were available, with the exception of 
old (and incorrect)  DW scattering calculations  for  \ion{Fe}{x}.

This paper is one in a series 
(see \citealt{delzanna_etal:04_fe_10}, hereafter Paper~I)
that aims to provide an 
assessment of  atomic data needed for the analysis of astrophysical spectra
by benchmarking them against all available experimental data.
The approach is observation-based, i.e. 
focuses  on the  brightest spectral lines that are observed
 in astrophysical  spectra.
The paper is organised as follows.
In Sect.~2, we give a brief review of previous 
observations we used for the benchmark. 
In Sect.~3 we present our results and in Sect.~4 we reach
our conclusions.

\section{Previous observations and line identifications}

The best soft X-ray spectra of the Sun in terms
of radiometric calibration  are currently 
provided by the SDO EVE spectrometers.
The SDO EVE spectra are calibrated with the use of sounding rockets 
that carry copies of the flight instruments, which in turn  are
carefully calibrated before and after each flight against a 
standard source.
On 2008 April 14, a prototype of the EVE instrument
was flown (hereafter PEVE). It provided an excellent EUV 
spectrum of the quiet Sun \citep{woods_etal:09,chamberlin_etal:09, delzanna_etal:10_cdscal}
that we use here for the benchmark.
 The  F10.7 radio flux on that day was only 
69. Indeed during the previous extended minimum 
the solar corona was very quiet \citep{delzanna_andretta:11}.
One drawback of the EVE spectra is the
low spectral resolution (about 1~\AA), hence the majority
of the lines are blended.

Very few solar soft X-ray high-resolution spectra exist, all being obtained
with rocket flights in the 1960's and observing the Sun as a star. 
As discussed in  \cite{delzanna_etal:10_cdscal, delzanna_andretta:11}, there is now 
good evidence that the basal quiet Sun irradiances in lines formed at or below
1 MK are relatively unchanged across solar cycles.
Also, that irradiances during solar minimum conditions
are similar for different cycles.
Hence, it is reasonable to compare irradiances
of the quiet Sun obtained over different periods. 
So we occasionally use the PEVE irradiances 
(obtained by fitting the original spectra) in conjunction
with the quiet Sun  irradiances of \cite{manson:72} [hereafter M72]
for the present benchmark.

M72 provided an excellent list of 
calibrated soft X-ray irradiances
 observed in quiet and active conditions in the 30--130~\AA\ range.
The quiet Sun spectrum was obtained on 1965 November 3,
when the F10.7 flux was 80.6.
The active Sun spectrum was obtained on 1967 August 8,
when the F10.7 flux was 143.4, i.e. when the Sun was relatively
active.
The spectral resolution was moderate, about 0.23~\AA\ (FWHM)
for the quiet Sun, and 0.16~\AA\ for the active Sun observation.

\cite{behring_etal:72}  [hereafter Be72]
published a line list from a spectrum obtained with a spectrograph built at 
the Goddard Space Flight Center and flown on an Aerobee 150 rocket flight
on 1969 May 16. On that day,   the F10.7 flux was 159.4,
i.e. the Sun was moderately active, as in the active Sun M72 observation.
The instrument observed the entire Sun in the  60-385 \AA\ region with 
high-resolution (0.06 \AA). 
To date, the Be72 spectrum is the best in terms of spectral resolution
and wavelength accuracy for the strongest lines in the Soft X-rays.
Unfortunately, only approximate intensities were provided.

\cite{malinovsky_heroux:73} [hereafter MH73]
 presented  an integrated-Sun spectrum 
covering the 50-300 \AA\ range with a medium resolution (0.25 \AA),
taken with  a grazing-incidence spectrometer  flown on a
rocket on 1969 April 4, when the F10.7 flux was 177.3, i.e.
when the Sun was `active'.
The spectrum was photometrically calibrated, and still,
quite surprisingly,  
represent the best available spectrum in the EUV (150--300) \AA\ range.
The tables provided by MH73  were not complete,
so we have scanned their spectra  to provide additional information.
At various wavelengths, the MH73 resolution was better than M72.
The spectra have been  wavelength and flux calibrated, matching the 
MH73 published intensities.

\begin{figure}[!htbp]
\centerline{\epsfig{file=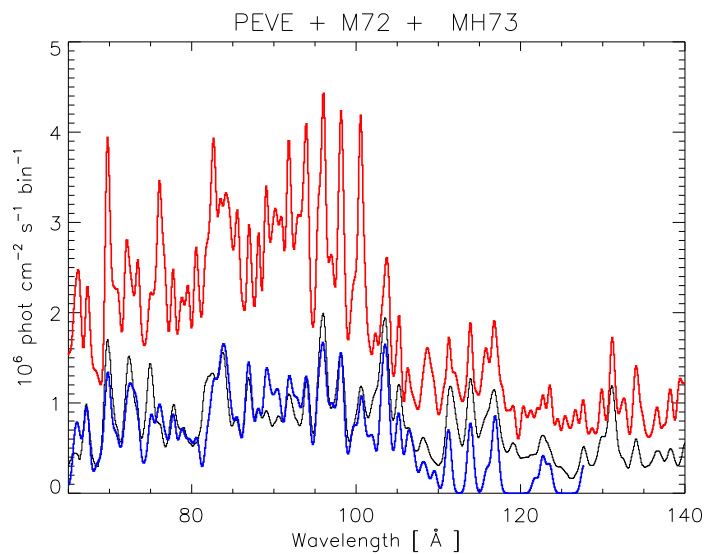, width=7.5cm,angle=90 }}
 \caption{A comparison between the soft X-ray irradiances
of PEVE (black thin line), M72 (thicker blue) and MH73
(thicker red).
}
\label{fig:irr}
\end{figure}

The MH73 irradiances were used by \cite{malinovsky_etal:80}
to benchmark their \ion{Fe}{x} atomic calculations.
The results were discouraging, with the ratios of the soft X-ray vs. the EUV
 lines being largely (by more than a factor of two) under-predicted by theory.
The actual atomic calculations were incorrect, however,
as pointed out in \cite{delzanna_etal:12_fe_10}.
Also, it turns out that the line irradiances were incorrect. 
A simple direct comparison of the published irradiances by 
MH73 and M72 clearly shows a discrepancy of about a factor of two
at various wavelengths.
Various comparisons with the quiet Sun PEVE spectrum have been
done, by taking into account the differences in spectral 
resolution. It is clear that the M72 has calibrated irradiances in excellent
agreement with the PEVE ones in the 60--100~\AA\  region, while the 
MH73 are largely overestimated, as shown in  Fig.~\ref{fig:irr}.
The M72 irradiances have been  obtained by convolving the published intensities
and putting them onto the PEVE resolution.
The large difference in the MH73 irradiances have nothing to do with the
fact that the Sun was more active, because they are present even in 
 cool lines, which have similar irradiances independently of the solar conditions.
We have therefore recalibrated the MH73 spectrum to agree with the PEVE one.
Obviously, in various spectral regions where `hot' lines are present,
some disagreement is present. A few of these recalibrated MH73
line irradiances are used for the present benchmark.

The M72 irradiances above 100~\AA\ are slowly decreasing when 
compared to the PEVE ones, an indication
of an incorrect calibration towards the longer wavelengths.
We have therefore also recalibrated the M72 quiet Sun spectrum 
above 100~\AA.

% Fig.~\ref{fig:irr}

\cite{acton_etal:85} [hereafter A85] published a high-quality solar spectrum 
recorded on photographic film %with an exposure of 145s 
during a rocket flight,
2 minutes after the GOES X-ray peak emission of an M1-class flare.
The spectrum was calibrated, and provided accurate line intensities,
although the sensitivity dropped above 77~\AA.
The spectral resolution was excellent, clearly resolving 
lines only 0.04~\AA\ apart. %Around 60~\AA\,  wavelengths 

The identification of the iron  soft X-rays $n=4 \to n=3$ transitions 
 started with the  pioneering (and to date best)  work by
{Edl{\' e}n} in the 1930's (see e.g. \citealt{edlen:37b} on \ion{Fe}{x}). 
Edl{\' e}n work was extended to the 
iron 3s$^2$3p$^2$~4$l$ ($l=$s,p,d,f) levels by
 the fundamental  laboratory work of \cite{fawcett_etal:72} [hereafter Fawcett].
It is important to keep in mind that only lines with
strong oscillator strengths were identified, that 
some identifications were tentative, and that a large
number of lines in the  spectra were left unidentified. 
We have re-analysed some of Fawcett's plates as part of 
a larger project to sort out the identifications in the 
 soft X-ray spectrum. They have been used here for the 
benchmark. Each plate has been scanned, average spectra obtained and 
wavelength-calibrated.

\section{Summary of the main results}

In order to assess how well experimental intensities compare 
with the predicted ones, we use the 
`emissivity ratio curves', introduced in Paper~I.
These curves are obtained by dividing the observed intensities
of the lines with their predicted emissivity as a function 
of the electron density (or temperature), 
calculated at a fixed temperature (or density),
and normalised to 1. 
The crossing (or small spreading around 1) of the curves indicates
agreement between observed and predicted line intensities.

The present benchmark is aimed at identifying the main 
transitions in some of the iron ions for which we have calculated
new atomic data, however it was necessary to benchmark also
a few other ions, to assess blending in the iron lines.

\subsection{\ion{Mg}{ix}}

The atomic data for  \ion{Mg}{ix} as calculated by 
\cite{delzanna_etal:08_mg_9} with the R-matrix method
have been used here. These APAP data are available  within CHIANTI
version 7 \citep{dere_etal:97, landi_etal:11_chianti_v7}.
The identifications are due to \cite{soderqvist:44_mg_9}.

\begin{figure}[!htbp]
\centerline{\epsfig{file=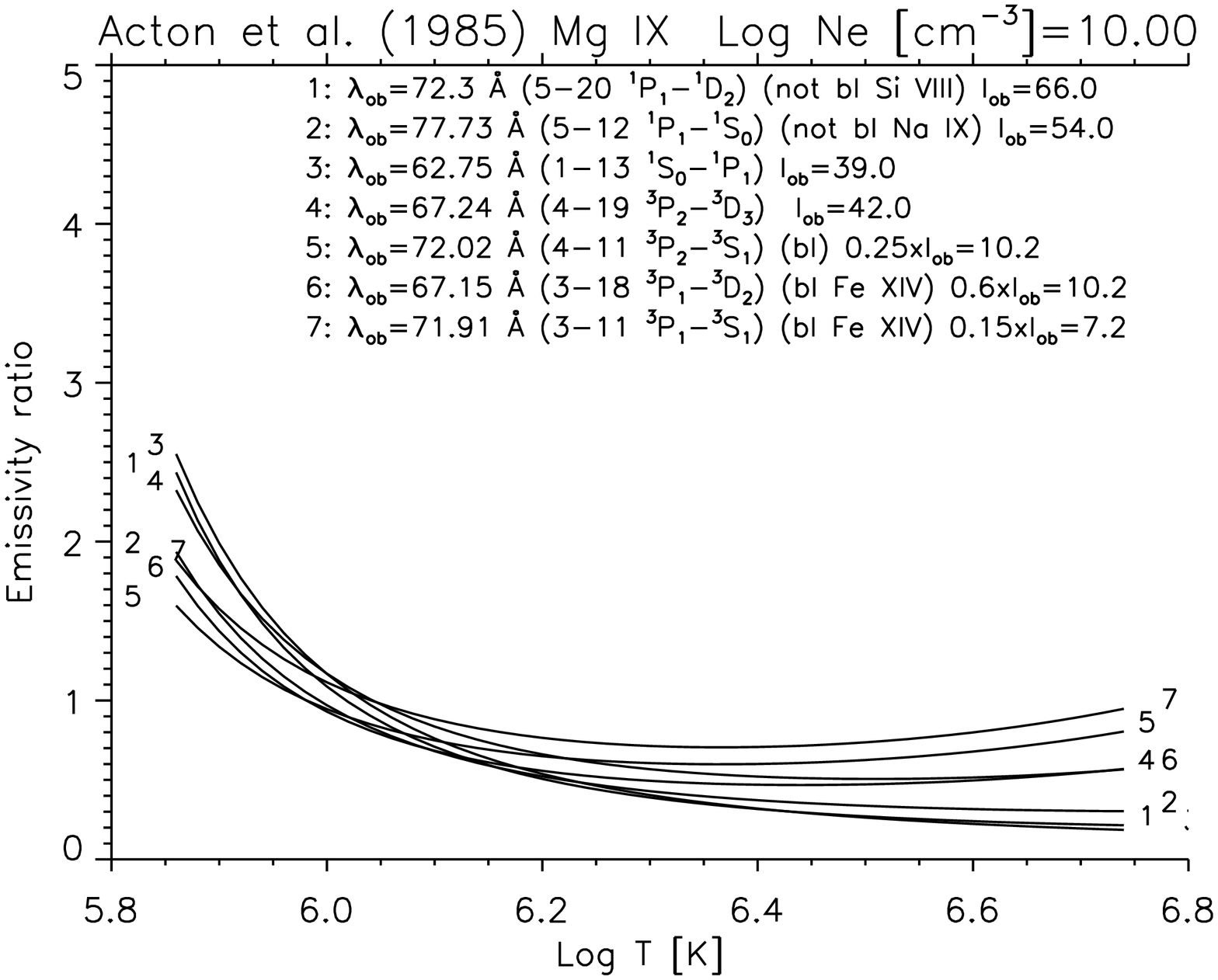, width=6.5cm,angle=90 }}
\centerline{\epsfig{file=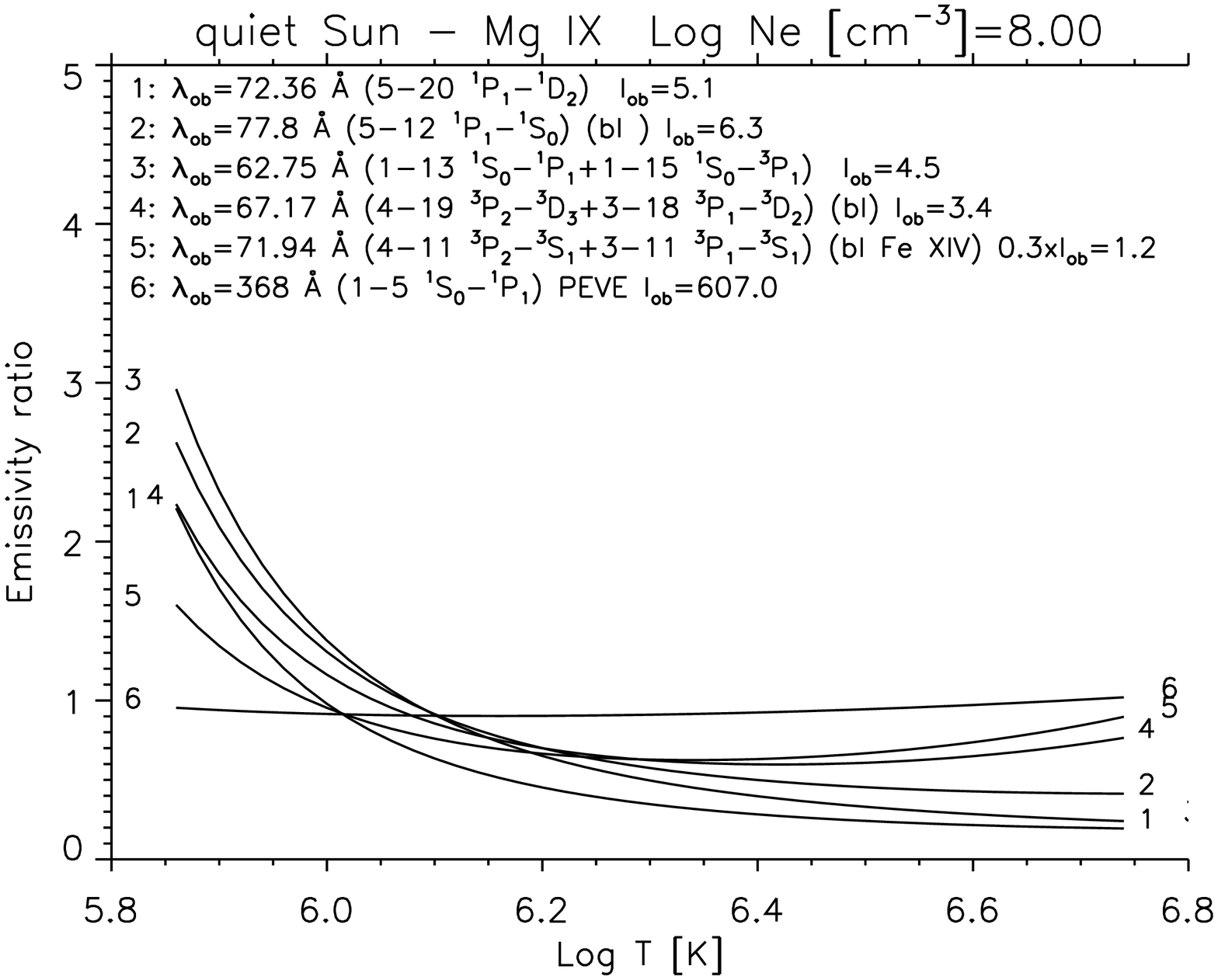, width=6.5cm,angle=90 }}
 \caption{Emissivity ratio  curves  relative to  the 
main \ion{Mg}{ix} lines and the solar flare (A85) and
quiet Sun  (M72, PEVE) observations.
$\lambda_{\rm ob}$ indicates the observed wavelength (\AA),
$I_{\rm ob}$ indicates the observed intensity, sometimes reduced
by the amount indicated. 
(bl) indicates the presence of a blend. }
 \label{fig:mg_9_A85}
\end{figure}
% Fig.~\ref{fig:mg_9_A85}

A85 identified a relatively large number of \ion{Mg}{ix} in the 
solar flare spectrum, however the benchmark has shown that a 
large number of those identifications are not correct.
Fig.~\ref{fig:mg_9_A85} (top) shows the emissivity ratio  curves  relative to  the 
main \ion{Mg}{ix} lines and the A85 observation.
The curves are plotted as a function of temperature because they have 
very little density sensitivity for the solar corona.

The 71.91~\AA\ line cannot be due to  \ion{Mg}{ix} as reported by A85.
By assuming that the strongest \ion{Mg}{ix}  transition at 72.30~\AA\ is
unblended, the 3--11  2s 2p $^3$P$_{1}$--2s 3s $^3$S$_{1}$ at 71.90~\AA\ should only 
account for about 15\% of the observed intensity, as shown in Fig.~\ref{fig:mg_9_A85}.
Similarly, the 4--11 2s 2p $^3$P$_{2}$--2s 3s $^3$S$_{1}$ transition can account
only about 25\% of the observed intensity at 72.02~\AA\ by A85.
Many of the \ion{Mg}{ix} are blended at the M72 resolution, however the 
two strongest lines, the 72.30 and 77.73~\AA\ lines, do not appear
to be blended with \ion{Si}{viii} and \ion{Na}{ix} as listed in A85.

Excellent agreement between the M72 quiet Sun irradiances and the PEVE 
irradiance of the 
resonance 368~\AA\ line is found (Fig.~\ref{fig:mg_9_A85} below),
for  a very reasonable electron temperature around 1~MK.
The PEVE measurement has been corrected
for the  \ion{Mg}{vii} contribution, estimated from  the SOHO/CDS
irradiances where the lines are resolved \citep{delzanna_andretta:11}.

\subsection{\ion{Mg}{x}}

\begin{figure}[!htbp]
\centerline{
\epsfig{file=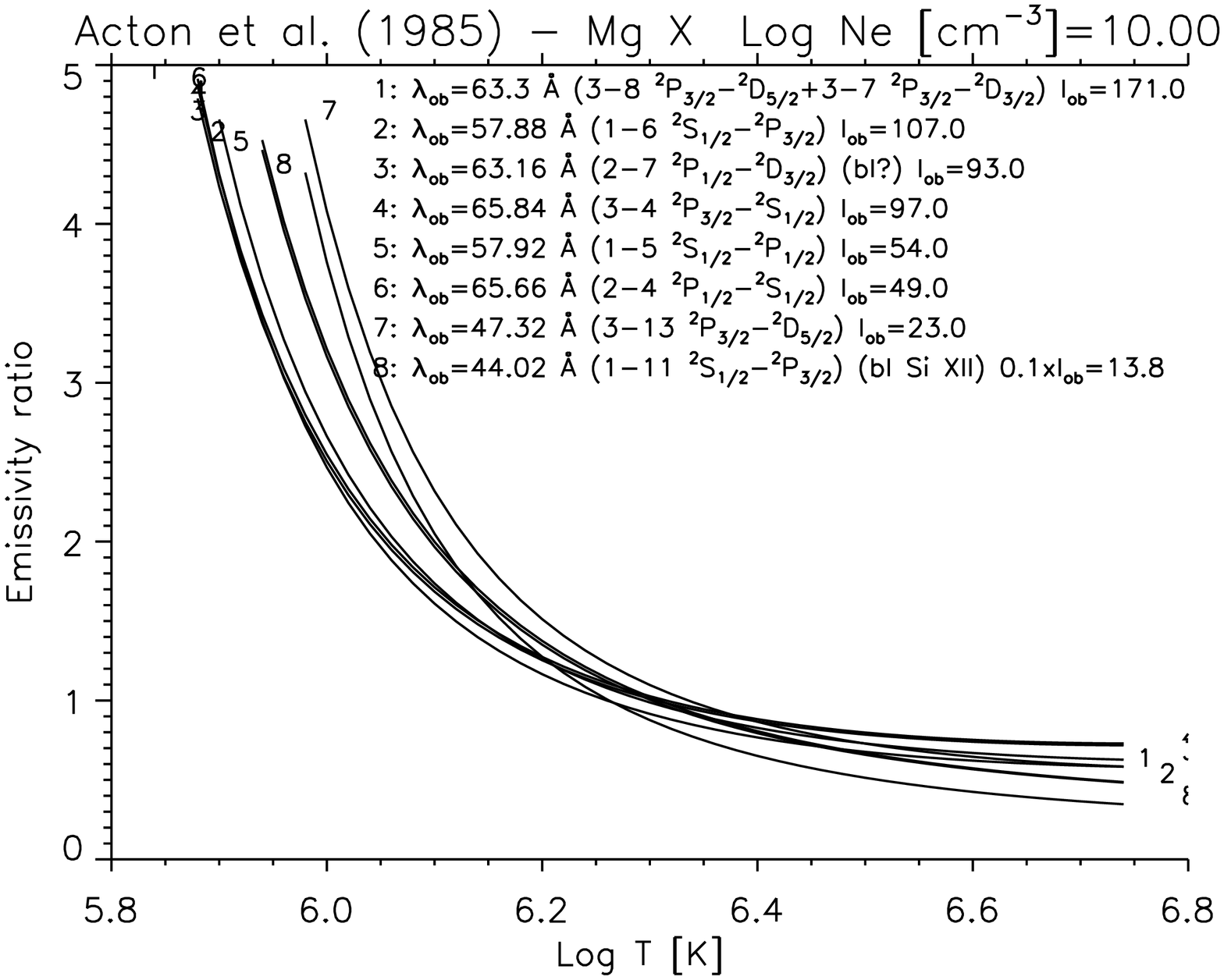, width=6.5cm,angle=90 }
}
 \caption{Emissivity ratio  curves  relative to  the 
main \ion{Mg}{x} lines and the A85 solar flare observation.}
 \label{fig:mg_10_A85}
\end{figure}
% Fig.~\ref{fig:mg_10_A85}

The excitation rates  for  \ion{Mg}{x} as calculated by 
\cite{zhang_etal:90} and available within CHIANTI have been used here.
The identifications are due to \cite{feldman_etal:70_li-like}.
The benchmark of this simple ion is straightforward. 
The identifications provided by A85 are confirmed, and excellent agreement 
between observed and predicted intensities is found, as 
shown in  Fig.~\ref{fig:mg_10_A85}. 
The curves are plotted as a function of temperature because they have 
no density sensitivity for the solar corona. 
Agreement within a few percent is obtained by assuming an isothermal 
temperature of log $T$[K]=6.2.
Only the weaker 1--11 44.05~\AA\ line
is blended with a stronger \ion{Si}{x}.

\begin{figure}[!htbp]
\centerline{\epsfig{file=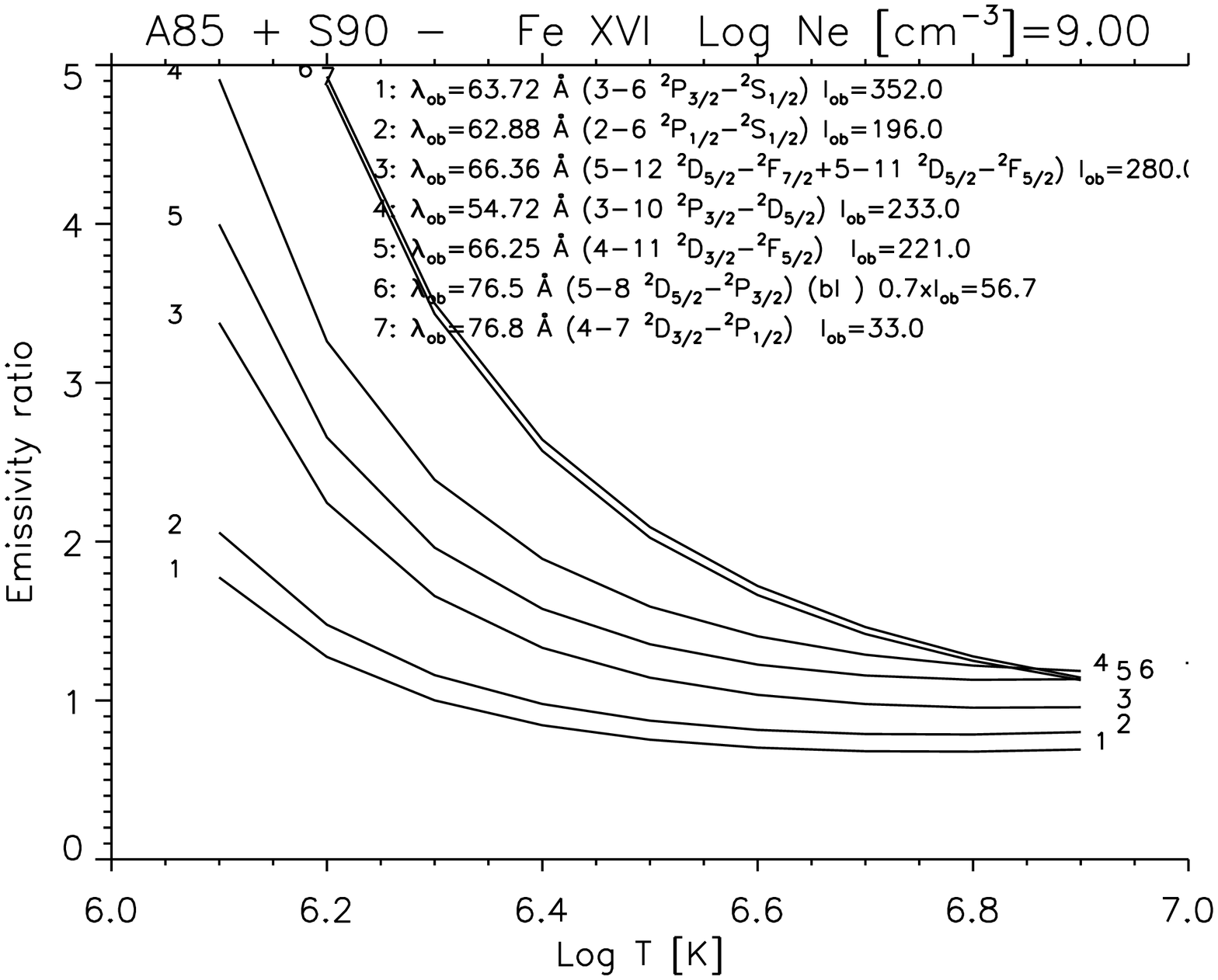, width=6.5cm,angle=90 }}
\centerline{\epsfig{file=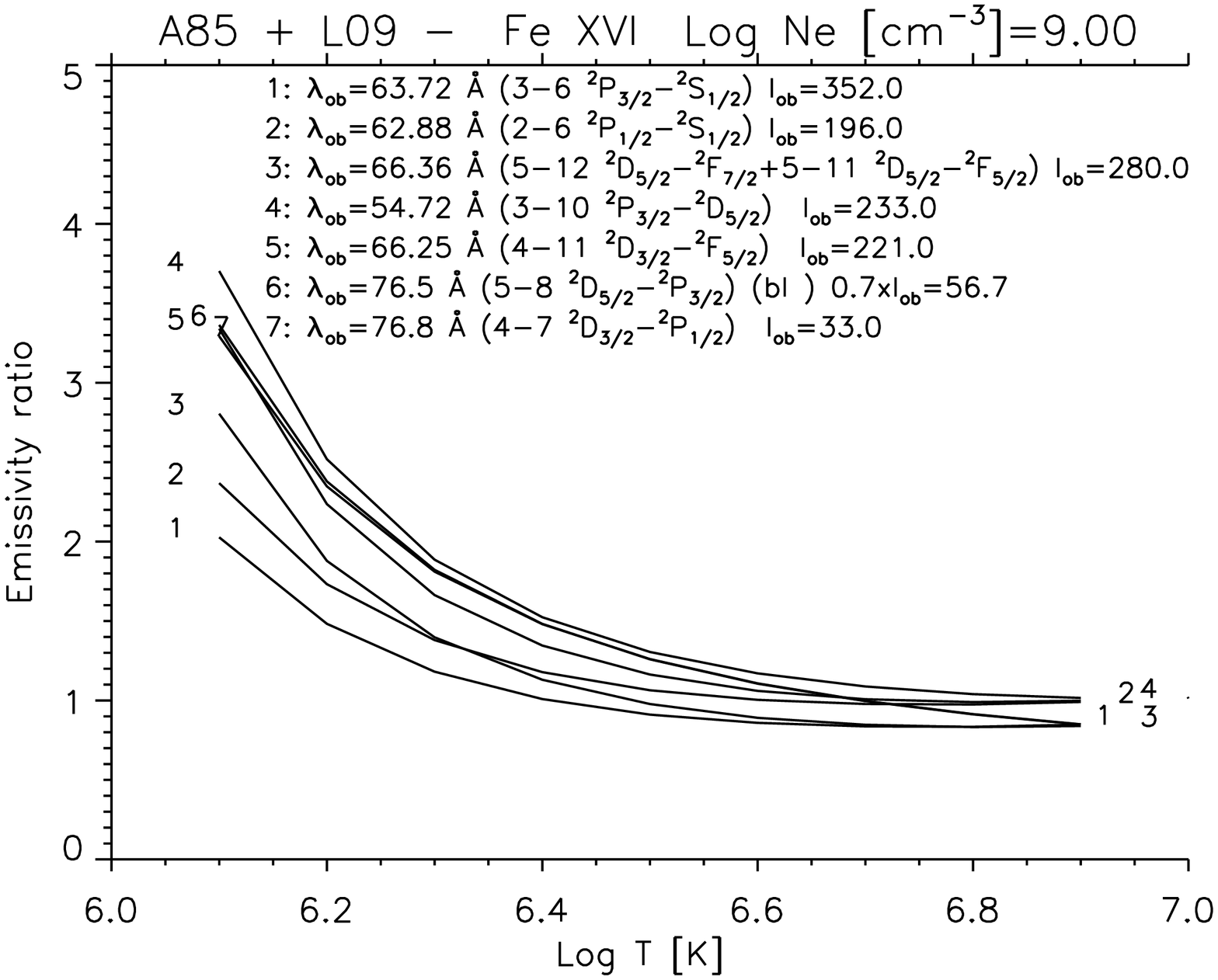, width=6.5cm,angle=90 }}
\caption{Emissivity ratio  curves  relative to  the 
main \ion{Fe}{xvi} lines and the A85 solar flare observation.
Top: using the \cite{sampson_et_al:90}  atomic data.
Bottom: using the APAP  data by \cite{liang_etal:09_na-like}.}
 \label{fig:fe_16}
\end{figure}
% Fig.~\ref{fig:fe_16}

\subsection{\ion{Fe}{xvi}}

The identifications of the \ion{Fe}{xvi} lines are due to 
\cite{edlen:36_na-like}.
The benchmark for this ion is straightforward. The A85 spectrum is excellent
for benchmarking the main lines from this ion, because these lines 
are very strong and well resolved.

\cite{sampson_et_al:90} performed relativistic DW calculations for this ion,
and the data are available within CHIANTI. 
\cite{cornille_etal:97} later performed a similar DW calculation, and 
pointed out the possible use of the  \ion{Fe}{xvi} lines to measure 
electron temperatures. However, very large values (above log$T$[K]=6.7) were obtained
for the A85 observation.
Various R-matrix  calculations have later been done. 
For example, \cite{aggarwal_keenan:06} perfomed a calculation with the 
Dirac Atomic R-matrix Code (DARC).  
These data were used by \cite{keenan_etal:07} to show that reasonable 
agreement for the A85 data was present, although they did not discuss the 
temperature sensitivity of these lines.
To show the large differences between DW and  R-matrix  calculations for this ion,
we plot in  Fig.~\ref{fig:fe_16} the emissivity ratio curves obtained with the
DW \cite{sampson_et_al:90} data and the latest  R-matrix  calculations
(within APAP) by \cite{liang_etal:09_na-like}.
The  curves in Fig.~\ref{fig:fe_16} are plotted  as a function of temperature 
because they have no density sensitivity for the solar corona. 
The large discrepancies and high temperatures are obvious when the DW data 
are considered, as  Fig.~\ref{fig:fe_16} (top) shows.
On the other hand, relatively good agreement is obtained with the R-matrix  calculations
(Fig.~\ref{fig:fe_16} bottom).
No significant temperature sensitivity is present.

\subsection{\ion{Fe}{xv}}

The  first identifications of the \ion{Fe}{xv} lines are due to 
\cite{edlen:36_mg-like}.  Fawcett identified several new transitions.
Later, \cite{cowan_widing:73} revised a few of Fawcett's identifications and
suggested a few tentative ones. Aside from \cite{edlen:36_mg-like}
and Fawcett, accurate wavelengths are given by \cite{kink_etal:97_fe_15}, where
a list of lines observed in 
laboratory spectra along the sequence is provided.

Various calculations for this ion exist in the literature.
\cite{bhatia_etal:97} performed a scattering calculation for this ion
complementing a DW run, and compared predicted line intensities
with those observed by A85. 
As in the  \ion{Fe}{xvi} case, large electron temperatures
 (above log$T$[K]=6.7) were obtained.
\cite{keenan_etal:06} used the \cite{aggarwal_etal:03} R-matrix calculations 
to find relatively good agreement between predicted and observed A85 intensities
at a much lower temperature (log$T$[K]=6.3).

Here, we use the atomic data available within CHIANTI v.7.
The atomic data for  the main $n=4$ levels  are
from the R-matrix calculations of 
\cite{berrington_etal:05_fe_15},  while those for the remaining  $n=4$ levels are 
from the DW calculations of \cite{landi:2011_fe_15}.
Table~\ref{tab:fe_15} lists the 
 relative intensities of the brightest soft X-ray lines in \ion{Fe}{xv},
at two densities, typical of the quiet solar corona and of laboratory spectra.

\begin{figure}[!htbp]
\centerline{\epsfig{file=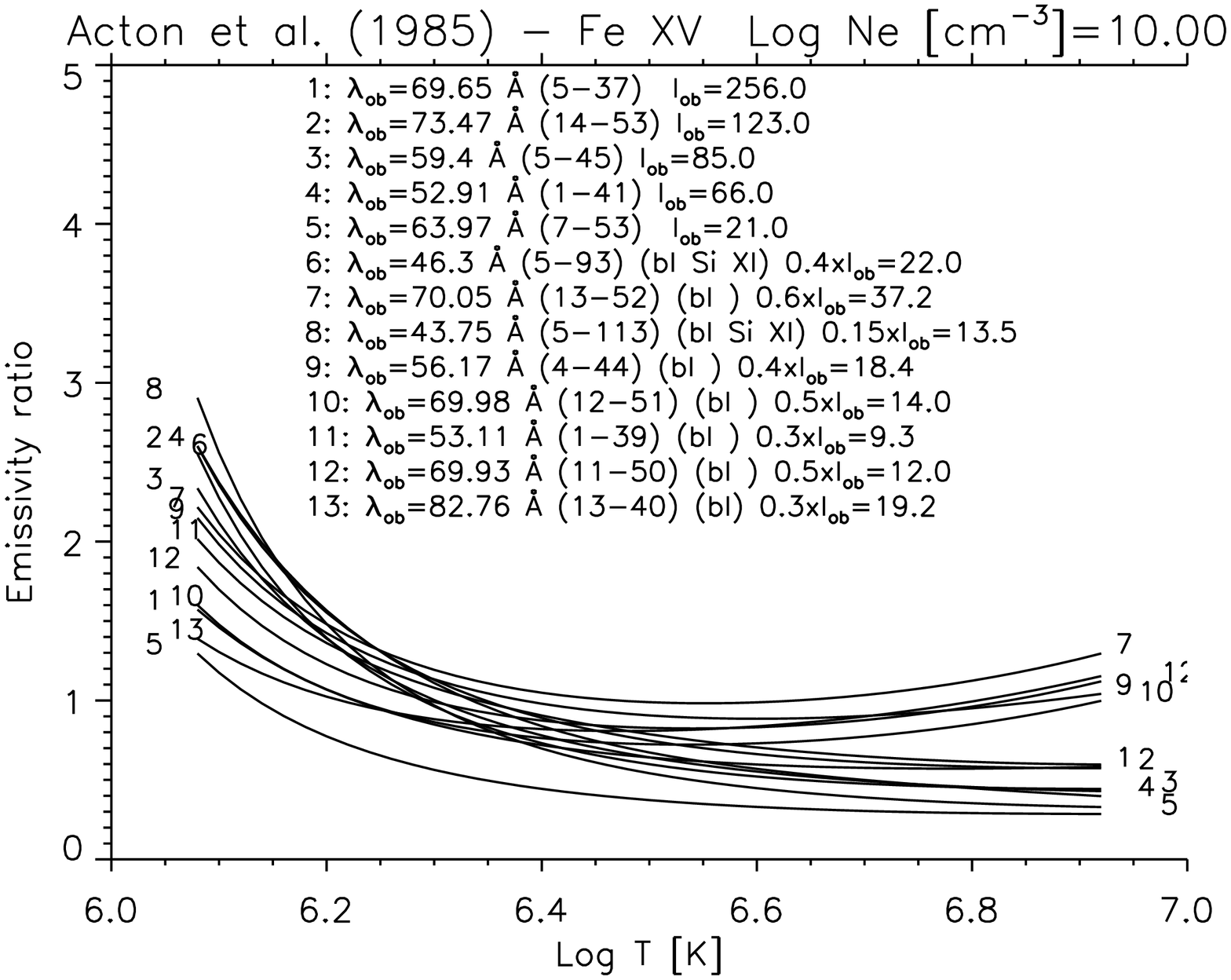, width=6.5cm,angle=90 }}
\caption{Emissivity ratio  curves  relative to  the 
main \ion{Fe}{xv} lines and the A85 solar flare observation.}
 \label{fig:fe_15}
\end{figure}
% Fig.~\ref{fig:fe_15}

\def\baselinestretch{1.}

\begin{table*}[!htbp]
\caption{The relative intensities of the brightest soft X-ray lines in \ion{Fe}{xv}.}
\begin{center}
%\scriptsize
\footnotesize
\begin{tabular}[c]{@{}llcrccrccccl@{}}
\hline\hline\noalign{\smallskip}  
 $i$--$j$ & Levels   &  $Int$ &  $Int$ & $gf$ &  A$_{ji}$(s$^{-1}$) &   $\lambda_{\rm exp}$(\AA) & $\lambda_{\rm th}$(\AA)   & New \\   
  &  & 1.0$\times$10$^{8}$ & 1.0$\times$10$^{19}$ &  &    &  \\ 
\noalign{\smallskip}\hline\noalign{\smallskip}                                   %inserts single line

5--37 & 3s 3p $^1$P$_{1}$--3s 4s $^1$S$_{0}$ & 1.0 & 5.1$\times$10$^{-2}$ & 0.16 & 2.2$\times$10$^{11}$ & 69.66 (? CW73) & 69.93 (0.3) & 69.661 (K97)  \\ 
14--53 & 3s 3d $^1$D$_{2}$--3s 4f $^1$F$_{3}$ & 0.36 & 2.4$\times$10$^{-2}$ & 3.45 & 6.0$\times$10$^{11}$ & 73.50 (CW73) & 73.82 (0.3) &   \\ 
5--45 & 3s 3p $^1$P$_{1}$--3s 4d $^1$D$_{2}$ & 0.28 & 1.8$\times$10$^{-2}$ & 0.64 & 2.4$\times$10$^{11}$ & 59.404 (F72) & 59.63 (0.2) &   \\ 
1--41 & 3s$^2$ $^1$S$_{0}$--3s 4p $^1$P$_{1}$ & 0.20 & 1.4$\times$10$^{-2}$ & 0.29 & 2.3$\times$10$^{11}$ & 52.911 (E36) & 52.96 (0.1) &   \\ 
7--53 & 3p$^2$ $^1$D$_{2}$--3s 4f $^1$F$_{3}$ & 0.12 & 8.1$\times$10$^{-3}$ & 0.89 & 2.1$\times$10$^{11}$ & 63.96 (? CW73) & 63.95 (-0.0) & 63.961 (K97)  \\ 
5--93 & 3s 3p $^1$P$_{1}$--3s 5s $^1$S$_{0}$ & 7.6$\times$10$^{-2}$ & 3.9$\times$10$^{-3}$ & 4.5$\times$10$^{-2}$ & 1.4$\times$10$^{11}$ &  -  & 46.28 & ? 46.30 (bl) \\ 
13--52 & 3s 3d $^3$D$_{3}$--3s 4f $^3$F$_{4}$ & 6.2$\times$10$^{-2}$ & 8.0$\times$10$^{-2}$ & 5.81 & 8.7$\times$10$^{11}$ & 70.054  (E36) & 70.17 (0.1) &   \\ 
5--113 & 3s 3p $^1$P$_{1}$--3s 5d $^1$D$_{2}$ & 4.6$\times$10$^{-2}$ & 3.0$\times$10$^{-3}$ & 0.27 & 1.8$\times$10$^{11}$ &  -  & 43.78 & ? 43.75 (bl)  \\ 
12--51 & 3s 3d $^3$D$_{2}$--3s 4f $^3$F$_{3}$ & 4.2$\times$10$^{-2}$ & 5.5$\times$10$^{-2}$ & 4.01 & 7.8$\times$10$^{11}$ & 69.987  (E36) & 70.11 (0.1) &   \\ 
11--50 & 3s 3d $^3$D$_{1}$--3s 4f $^3$F$_{2}$ & 3.3$\times$10$^{-2}$ & 3.7$\times$10$^{-2}$ & 2.70 & 7.3$\times$10$^{11}$ & 69.945  (E36) & 70.07 (0.1) &   \\ 
4--44 & 3s 3p $^3$P$_{2}$--3s 4d $^3$D$_{3}$ & 3.2$\times$10$^{-2}$ & 4.3$\times$10$^{-2}$ & 1.43 & 4.3$\times$10$^{11}$ & 56.200  (E36) & 56.22 (0.0) &   \\ 
7--41 & 3p$^2$ $^1$D$_{2}$--3s 4p $^1$P$_{1}$ & 3.1$\times$10$^{-2}$ & 2.2$\times$10$^{-3}$ & 9.3$\times$10$^{-2}$ & 3.6$\times$10$^{10}$ & 75.167  (E36) & 75.31 (0.1) &   \\ 

% wrong wavelength in CHIANTI 49.490
14--117 & 3s 3d $^1$D$_{2}$--3s 5f $^1$F$_{3}$ & 2.9$\times$10$^{-2}$ & 2.2$\times$10$^{-3}$ & 0.60 & 2.1$\times$10$^{11}$ & -  & 52.35 & ? 52.36  \\ 

 13--40 & 3s 3d $^3$D$_{3}$--3s 4p $^3$P$_{2}$ & 2.8$\times$10$^{-2}$ & 8.0$\times$10$^{-2}$ & 0.30 & 5.7$\times$10$^{10}$ &  -  & 82.98 & ? 82.750 (K97) \\ 

%9--53 & 3p$^2$ $^3$P$_{2}$--3s 4f $^1$F$_{3}$ & 2.7$\times$10$^{-2}$ & 1.7$\times$10$^{-3}$ & 0.20 & 4.5$\times$10$^{10}$ & 64.878 & 64.91 (0.0) &   \\ 

1--39 & 3s$^2$ $^1$S$_{0}$--3s 4p $^3$P$_{1}$ & 2.6$\times$10$^{-2}$ & 3.4$\times$10$^{-2}$ & 0.12 & 9.2$\times$10$^{10}$ &  -  & 53.17 & ? 53.11 (bl)  \\ 

\noalign{\smallskip}\hline\noalign{\smallskip}                                   %inserts single line
\end{tabular}
\normalsize
\tablefoot{The relative line intensities  (photons)  $Int=N_{j} A_{ji}/N_{\rm e}$ 
were calculated at log N$_{\rm e}$ [cm$^{-3}$]=8,19 and log $T$$_{\rm e}$ [K]= 6.3.
The lines are ordered with decreasing intensity.
The oscillator strengths and transition probabilities are shown.
The last three columns show the experimental 
wavelengths $\lambda_{\rm exp}$(\AA), when known, the
target wavelengths  $\lambda_{\rm th}$(\AA), with their difference
in parenthesis, and finally the new wavelengths proposed here.
We also add next to the experimental wavelength the reference
(E36: \citealt{edlen:36_mg-like}; F72: \citealt{fawcett_etal:72};
CW73:  \citealt{cowan_widing:73}).
A question mark indicates a tentative identification.
}
\end{center}
\label{tab:fe_15}
\end{table*}
% Table~\ref{tab:fe_15}

%\def\baselinestretch{1.5}

The A85  solar flare  spectrum is excellent
for benchmarking the main lines from this ion, because these lines 
are  strong and well resolved.
The emissivity ratio  curves  relative to  the 
main \ion{Fe}{xv} lines and the A85 observation are shown in Fig.~\ref{fig:fe_15}.
As \cite{keenan_etal:06} pointed out, some line ratios are sensitive to 
the electron density, while others to temperature.
Among the lines  considered here, lines no.7 and 9 (70.05 and 56.17~\AA) are the 
only ones  sensitive to density, so the emissivity curves are plotted as a function
of temperature, for a density appropiate for the A85 flare.

Excellent agreement is found for the four strongest lines, while the others
appear blended. 
A significant discrepancy is present for the 63.97~\AA\ line (no.5 in  Fig.~\ref{fig:fe_15}),
indicating a possible problem with the 
\cite{landi:2011_fe_15} data.
Overall, the results are slightly different than those presented by 
\cite{keenan_etal:06}.
The strongest transition, the 5--37 3s 3p $^1$P$_{1}$--3s 4s $^1$S$_{0}$,
was only tentatively identified by \cite{cowan_widing:73}, based on the fact that
 the 69.66~\AA\ line  becomes 
one of the strongest lines in the soft X-rays in solar flare conditions.
The identification was confirmed by \cite{bhatia_etal:97}.
\cite{kink_etal:97_fe_15} provides a wavelength of 69.661~\AA.
We also confirm the other tentative identification  by \cite{cowan_widing:73}
for the line at 63.96~\AA. For the other strongest lines, we confirm the 
identifications by \cite{edlen:36_mg-like} and Fawcett.
Table~\ref{tab:fe_15} also provides several tentative identifications
proposed here. \cite{keenan_etal:06} proposed the identification of the 
82.76~\AA\ line as the \ion{Fe}{xv} 13--40 transition, however the 
CHIANTI model suggests that only 30\% of the line is due to \ion{Fe}{xv}.

\subsection{\ion{Fe}{xiv}}

\def\baselinestretch{1.}

\begin{table*}[!htbp]
\caption{The relative intensities of the brightest soft X-ray lines in \ion{Fe}{xiv}.}
\begin{center}
%\scriptsize
\footnotesize
\begin{tabular}[c]{@{}llcrccrccccl@{}}
\hline\hline\noalign{\smallskip}  
% fe_14_emiss_new.tex
% ss/atom/fe_14/liang/IC_so_oe
% Line details. The relative intensities  (photons)  $Int=N_{j} A_{ji}/N_{\rm e}$ are normalised to 3.2e-09} \label{tab:lines} \\
 $i$--$j$ & Levels   &  $Int$ &  $Int$ & $gf$ &  A$_{ji}$(s$^{-1}$) &   $\lambda_{\rm exp}$(\AA) & $\lambda_{\rm th}$(\AA)   & New \\   
  &  & 1.0$\times$10$^{8}$ & 1.0$\times$10$^{19}$ &  &    &  \\ 
\noalign{\smallskip}\hline\noalign{\smallskip}                                   %inserts single line

6--136 & 3s 3p$^2$ $^2$D$_{3/2}$--3s 3p 4s $^2$P$_{1/2}$ & 1.0 & 2.0$\times$10$^{-2}$ & 0.26 & 1.7$\times$10$^{11}$ &  -  & 71.37 &  71.919  \\ 
11--146 & 3s$^2$ 3d $^2$D$_{3/2}$--3s$^2$ 4f $^2$F$_{5/2}$ & 0.67 & 9.7$\times$10$^{-2}$ & 3.12 & 6.0$\times$10$^{11}$ & 76.022 & 76.04 (0.0) &   \\ 

11--122 & 3s$^2$ 3d $^2$D$_{3/2}$--3s$^2$ 4p $^2$P$_{1/2}$ & 0.44 & 6.5$\times$10$^{-3}$ & 9.6$\times$10$^{-2}$ & 3.9$\times$10$^{10}$ & 91.273 & 93.96 (2.7) & 93.618  (bl)\\
6--122 & 3s 3p$^2$ $^2$D$_{3/2}$--3s$^2$ 4p $^2$P$_{1/2}$ & 0.36 & 5.4$\times$10$^{-3}$ & 5.9$\times$10$^{-2}$ & 3.2$\times$10$^{10}$ & 78.765 & 80.25 (1.5) & 80.50  (bl)  \\
 
1--137 & 3s$^2$ 3p $^2$P$_{1/2}$--3s$^2$ 4d $^2$D$_{3/2}$ & 0.33 & 4.4$\times$10$^{-2}$ & 0.55 & 2.6$\times$10$^{11}$ & 58.963 & 58.80 (-0.2) &  ? 58.92 \\ 

8--136 & 3s 3p$^2$ $^2$S$_{1/2}$--3s 3p 4s $^2$P$_{1/2}$ & 0.25 & 5.1$\times$10$^{-3}$ & 7.3$\times$10$^{-2}$ & 4.3$\times$10$^{10}$ &  -  & 75.08 &  75.469  (bl) \\ 

2--101 & 3s$^2$ 3p $^2$P$_{3/2}$--3s$^2$ 4s $^2$S$_{1/2}$ & 0.25 & 3.2$\times$10$^{-2}$ & 0.29 & 1.9$\times$10$^{11}$ & 70.613 & 70.56 (-0.1) &   \\ 
12--148 & 3s$^2$ 3d $^2$D$_{5/2}$--3s$^2$ 4f $^2$F$_{7/2}$ & 0.19 & 0.14 & 4.49 & 6.5$\times$10$^{11}$ & 76.151 & 76.16 (0.0) &   \\ 

4--179 & 3s 3p$^2$ $^4$P$_{3/2}$--3s 3p 4d $^4$D$_{5/2}$ & 0.13 & 5.8$\times$10$^{-2}$ & 1.09 & 3.5$\times$10$^{11}$ &  -  & 58.71 & ? 58.79  \\ 
5--184 & 3s 3p$^2$ $^4$P$_{5/2}$--3s 3p 4d $^4$D$_{7/2}$ & 0.12 & 0.10 & 1.87 & 4.4$\times$10$^{11}$ &  -  & 58.89                &  ? 58.96  \\ 

2--138 & 3s$^2$ 3p $^2$P$_{3/2}$--3s$^2$ 4d $^2$D$_{5/2}$ & 0.12 & 8.2$\times$10$^{-2}$ & 1.01 & 3.1$\times$10$^{11}$ & 59.579 & 59.39 (-0.2) &   \\ 
6--146 & 3s 3p$^2$ $^2$D$_{3/2}$--3s$^2$ 4f $^2$F$_{5/2}$ & 0.12 & 1.7$\times$10$^{-2}$ & 0.43 & 1.1$\times$10$^{11}$ & 67.141 & 66.81 (-0.3) &   \\ 
1--101 & 3s$^2$ 3p $^2$P$_{1/2}$--3s$^2$ 4s $^2$S$_{1/2}$ & 0.12 & 1.5$\times$10$^{-2}$ & 0.13 & 9.2$\times$10$^{10}$ & 69.685 & 69.66 (-0.0) &   \\ 

11--136 & 3s$^2$ 3d $^2$D$_{3/2}$--3s 3p 4s $^2$P$_{1/2}$ & 0.11 & 2.2$\times$10$^{-3}$ & 3.6$\times$10$^{-2}$ & 1.8$\times$10$^{10}$ &  -  & 82.01 &  82.23 \\ 
12--125 & 3s$^2$ 3d $^2$D$_{5/2}$--3s$^2$ 4p $^2$P$_{3/2}$ & 9.9$\times$10$^{-2}$ & 1.1$\times$10$^{-2}$ & 0.16 & 3.3$\times$10$^{10}$ & 91.008 & 93.50 (2.5) & 93.20   \\ 
7--125 & 3s 3p$^2$ $^2$D$_{5/2}$--3s$^2$ 4p $^2$P$_{3/2}$ & 8.4$\times$10$^{-2}$ & 9.8$\times$10$^{-3}$ & 0.10 & 2.8$\times$10$^{10}$ & 78.583 & 79.90 (1.3) & 80.21  \\ 

\noalign{\smallskip}

1--8 & 3s$^2$ 3p $^2$P$_{1/2}$--3s 3p$^2$ $^2$S$_{1/2}$ & 26. & 0.12 & 0.39 & 1.7$\times$10$^{10}$ & 274.203 & 272.03 (-2.2) &   \\

\noalign{\smallskip}\hline\noalign{\smallskip}                                   %inserts single line
\end{tabular}
\normalsize
\tablefoot{The relative line intensities  (photons)  $Int=N_{j} A_{ji}/N_{\rm e}$ 
were calculated at log N$_{\rm e}$ [cm$^{-3}$]=8,19 and log $T$$_{\rm e}$ [K]= 6.3.}
%The lines are ordered with decreasing intensity.
%The oscillator strengths and transition probabilities are shown.
%The last three columns show the experimental 
%wavelengths $\lambda_{\rm exp}$(\AA), when known, the
%target wavelengths  $\lambda_{\rm th}$(\AA), with their difference
%in parenthesis, and finally the new wavelengths proposed here.
%A question mark indicates a tentative identification.
%}
\end{center}
\label{tab:fe_14}
\end{table*}
% Table~\ref{tab:fe_14}

%\def\baselinestretch{1.5}

\begin{figure}[!htbp]
\centerline{\epsfig{file=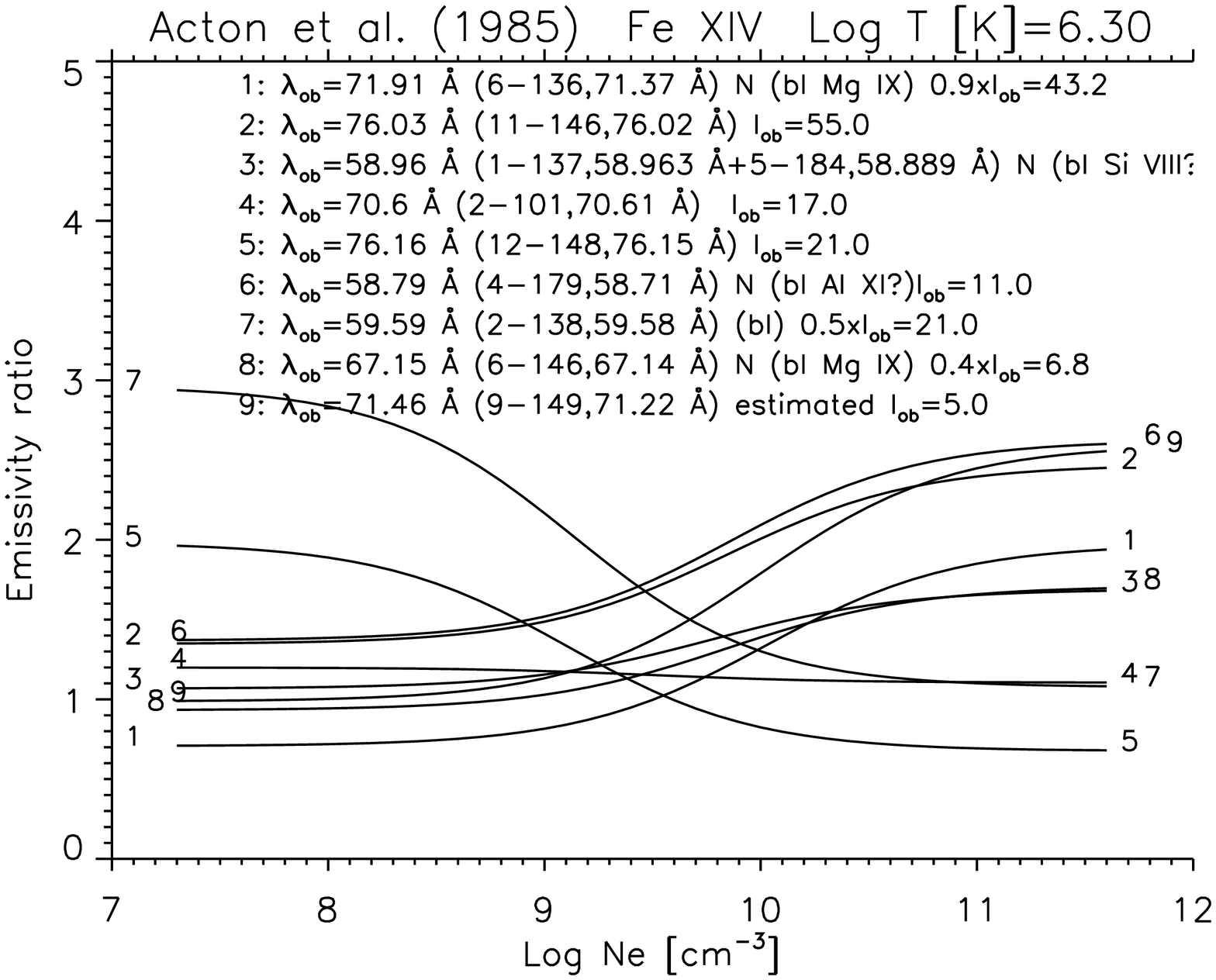, width=6.5cm,angle=90 }}
\centerline{\epsfig{file=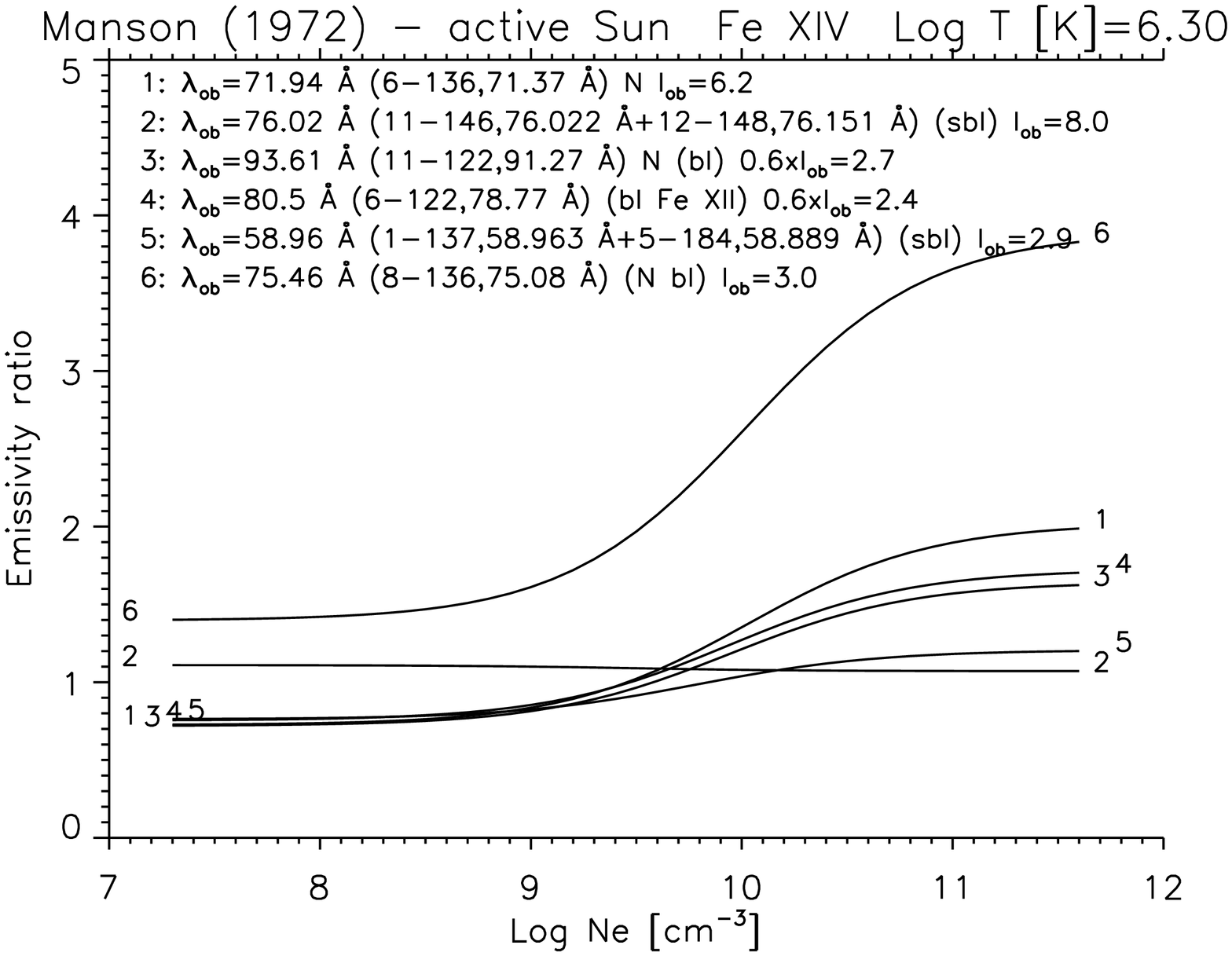, width=6.5cm,angle=90 }}
\caption{Emissivity ratio  curves  relative to  the 
main \ion{Fe}{xiv} lines. Top: A85 solar flare observation.
Bottom: M72 active sun observation.}
 \label{fig:fe_14}
\end{figure}
% Fig.~\ref{fig:fe_14}

The atomic data for  \ion{Fe}{xiv} have recently been calculated by 
\cite{liang_etal:10_fe_14} within the APAP network, and are used here.
Table~\ref{tab:fe_14} lists the 
 relative intensities of the brightest soft X-ray lines in \ion{Fe}{xiv},
at two densities, typical of the quiet solar corona and of laboratory spectra.
The identifications of the sof X-ray lines are from Fawcett.
Most of the lines identified by Fawcett are very close 
(within 0.3~\AA) to  the target wavelengths. 
For example,  the strongest decay from the 
3s$^2$ 4s is the 2--101  3s$^2$ 3p $^2$P$_{3/2}$--3s$^2$ 4s $^2$S$_{1/2}$ transition,
at a target wavelength of 70.56~\AA, and identified by Fawcett with the 
70.61~\AA\ line.
However, some are unidentified and some have large
departures, indicating likely misidentifications.

The strongest soft X-ray line,previously not identified, is the main decay (6--136) to the 
3s 3p$^2$ $^2$D$_{3/2}$ from the 3s 3p 4s $^2$P$_{1/2}$, which has a large population
due to a strong forbidden core-excited transition from the ground state.
The target wavelength for the 6--136 transition  is 71.37~\AA.  The only strong line around 71.37~\AA\ is the 
71.91~\AA\ line, previously incorrectly identified by A85 with a \ion{Mg}{ix} transition
as seen previously.
Be72 wavelength is 71.919~\AA.
There is a strong line in Fawcett's  C53 plate at 71.94~\AA. 
The A85 intensity of the 71.91~\AA\ line (corrected for a 10\% contribution from \ion{Mg}{ix} as discussed
previously) is in excellent agreement with that of the 70.61~\AA\ line, at 
log N$_{\rm e}$ [cm$^{-3}$]=9.8.
Good agreement is also found with the 12--148  3s$^2$ 3d $^2$D$_{5/2}$--3s$^2$ 4f $^2$F$_{7/2}$ 76.15~\AA\
line as  Fig.~\ref{fig:fe_14} shows.
The intensities measured by M72 for the active Sun also confirm the identification.
The second decay from the 3s 3p 4s $^2$P$_{1/2}$ is the  weaker (and blended)
8--136 transition, observed by M72 at 75.46~\AA. Be72 reports a wavelength of 75.469~\AA,
in excellent agreement with what predicted from the  wavelength of the 6--136 line (75.471~\AA).

\def\baselinestretch{1.}

\begin{table*}[!htbp]
\caption{The relative intensities of the brightest soft X-ray lines in \ion{Fe}{xiii}.}
\begin{center}
%\scriptsize
\footnotesize
\begin{tabular}[c]{@{}llcrccrccccl@{}}
\hline\hline\noalign{\smallskip}  

% File: fe_13_rm4+dw6_wp_8_19_6.25_emiss
 $i$--$j$ & Levels   &  $Int$ &  $Int$ & $gf$ &  A$_{ji}$(s$^{-1}$) &   $\lambda_{\rm exp}$(\AA) & $\lambda_{\rm th}$(\AA)   & New \\   
  &  & 1.0$\times$10$^{8}$ & 1.0$\times$10$^{19}$ &  &    &  \\ 
\noalign{\smallskip}\hline\noalign{\smallskip}                                   %inserts single line

7--331 & 3s 3p$^3$ $^3$D$_{1}$--3s 3p$^2$ 4s $^3$P$_{0}$ & 1.0 & 6.9$\times$10$^{-3}$ & 0.12 & 1.4$\times$10$^{11}$ &  -  & 75.71 &  76.507 \\ 

7--265 & 3s 3p$^3$ $^3$D$_{1}$--3s$^2$ 3p 4p $^3$P$_{0}$ & 0.70 & 2.1$\times$10$^{-3}$ & 3.2$\times$10$^{-2}$ & 3.0$\times$10$^{10}$ &  -  & 84.17 & 85.47 (bl)  \\ 

% Be72 82.425
20--409 & 3s$^2$ 3p 3d $^3$P$_{1}$--3s$^2$ 3p 4f $^3$F$_{2}$ & 0.53 & 2.6$\times$10$^{-2}$ & 1.22 & 2.4$\times$10$^{11}$ &  -  & 81.65 &   82.425  (bl) \\ 

20--265 & 3s$^2$ 3p 3d $^3$P$_{1}$--3s$^2$ 3p 4p $^3$P$_{0}$ & 0.50 & 1.5$\times$10$^{-3}$ & 3.5$\times$10$^{-2}$ & 2.1$\times$10$^{10}$ & - & 102.91& 103.928 (bl) \\
 
1--341 & 3s$^2$ 3p$^2$ $^3$P$_{0}$--3s$^2$ 3p 4d $^3$D$_{1}$ & 0.40 & 2.2$\times$10$^{-2}$ & 0.38 & 2.1$\times$10$^{11}$ & 62.353 & 61.74 (-0.6) &   \\ 

% Be72 83.221 
23--409 & 3s$^2$ 3p 3d $^3$D$_{1}$--3s$^2$ 3p 4f $^3$F$_{2}$ & 0.39 & 1.9$\times$10$^{-2}$ & 0.91 & 1.7$\times$10$^{11}$ &  -  & 82.43 &  83.221  \\

3--210 & 3s$^2$ 3p$^2$ $^3$P$_{2}$--3s$^2$ 3p 4s $^3$P$_{1}$ & 0.28 & 1.8$\times$10$^{-2}$ & 0.19 & 7.3$\times$10$^{10}$ & 75.892 & 75.05 (-0.8) &   \\ 

11--331 & 3s 3p$^3$ $^3$P$_{1}$--3s 3p$^2$ 4s $^3$P$_{0}$ & 0.22 & 1.5$\times$10$^{-3}$ & 2.9$\times$10$^{-2}$ & 3.1$\times$10$^{10}$ &  -  & 78.31 & 79.08 (bl)  \\ 

3--344 & 3s$^2$ 3p$^2$ $^3$P$_{2}$--3s$^2$ 3p 4d $^3$D$_{3}$ & 0.22 & 5.7$\times$10$^{-2}$ & 0.94 & 2.2$\times$10$^{11}$ & 62.975 & 62.33 (-0.6) &   \\ 

16--259 & 3s$^2$ 3p 3d $^3$F$_{3}$--3s$^2$ 3p 4p $^3$D$_{2}$ & 0.20 & 1.3$\times$10$^{-2}$ & 0.28 & 3.8$\times$10$^{10}$ &  98.523  & 97.82 &   \\ 

2--341 & 3s$^2$ 3p$^2$ $^3$P$_{1}$--3s$^2$ 3p 4d $^3$D$_{1}$ & 0.17 & 9.2$\times$10$^{-3}$ & 0.16 & 8.9$\times$10$^{10}$ & 62.717 & 62.08 (-0.6) & (bl)  \\

% 4--221 & 3s$^2$ 3p$^2$ $^1$D$_{2}$--3s$^2$ 3p 4s $^1$P$_{1}$ & 3.7$\times$10$^{-2}$ & 4.1$\times$10$^{-2}$ & 0.33 & 1.2$\times$10$^{11}$ & 76.117 & 75.41 (-0.7) &   \\ 
\noalign{\smallskip}\hline\noalign{\smallskip}                                   %inserts single line
\end{tabular}
\normalsize
\tablefoot{The relative line intensities  (photons)  $Int=N_{j} A_{ji}/N_{\rm e}$ 
were calculated at log N$_{\rm e}$ [cm$^{-3}$]=8,19 and log $T$$_{\rm e}$ [K]= 6.3
}
\end{center}
\label{tab:fe_13}
\end{table*}
% Table~\ref{tab:fe_13}

%\def\baselinestretch{1.5}

The strongest decays from the 3s 3p 4d 
(see 4--179 and 5--184 in Table)  are tentatively identified here 
with the lines observed by A85 at 58.79 and 58.96~\AA.
The two main decays from the 3s$^2$ 4d (1--137 and  2--138) were identified by \cite{edlen:36_mg-like}.
If the identifications are correct, the first would be a self-blend and the 
second severely blended in the A85 spectrum.

The two main decays (11-122 and 12-125) from the 3s$^2$ 4p $^2$P$_{1/2,3/2}$ levels  were identified
by Fawcett at 91.273 and 91.009~\AA\ respectively. The first
is predicted to be the third strongest \ion{Fe}{xiv} solar soft X-ray line.
In the M72 and MH73 spectra of the quiet Sun there are no strong lines
at this wavelength. 
Furthermore, the 91.273~\AA\ wavelength is at odds (2.7~\AA) with the predicted one. 
The identification is therefore incorrect. The only line that matches
well the predicted intensity and wavelength is the solar line at 
93.61~\AA, also observed in Fawcett's plate C53 at exactly the same 
 wavelength. 
Be72 lists a strong line at 93.618~\AA.
M72 clearly showed that this line becomes enhanced in 
active Sun conditions, which indicates that the line must be formed
around 3 MK, the average temperature of active region cores, which is another
argument in favor of the present identification as \ion{Fe}{xiv}.

This line is of particular importance for the SDO AIA 94~\AA\ band, 
as discussed below.
 Fig.~\ref{fig:fe_14} shows that about 60\% of the
intensity observed by M72 in the active Sun can be accounted by the \ion{Fe}{xiv}
11--122  3s$^2$ 3d $^2$D$_{3/2}$--3s$^2$ 4p $^2$P$_{1/2}$ 93.61~\AA\ line.
The decay to 3s 3p$^2$ $^2$D$_{3/2}$ (6--122 line) was identified by 
Fawcett with a line at 78.765~\AA. From the new wavelength of 93.61~\AA\
we obtain a wavelength of 80.50~\AA\ for the 6-122 line.
In both solar and laboratory plates there is a very strong broad 
line around 80.50~\AA, partly due to  \ion{Fe}{xii} (see below).
The weaker decays from level 125 are identified with lines
at 93.20 and 80.21~\AA.

\subsection{\ion{Fe}{xiii}}

\begin{figure}[!htbp]
\centerline{\epsfig{file=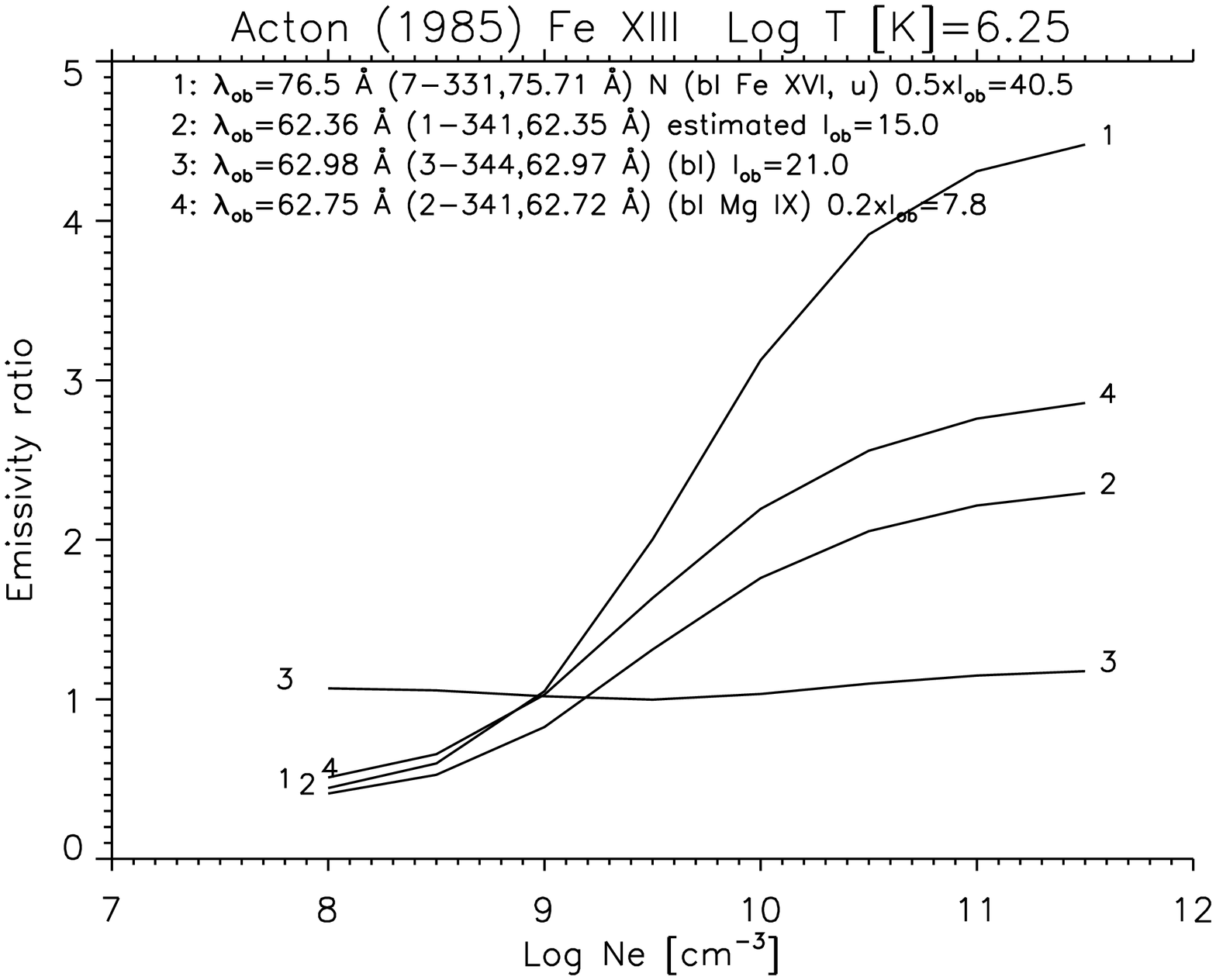, width=6.5cm,angle=90 }}
\centerline{\epsfig{file=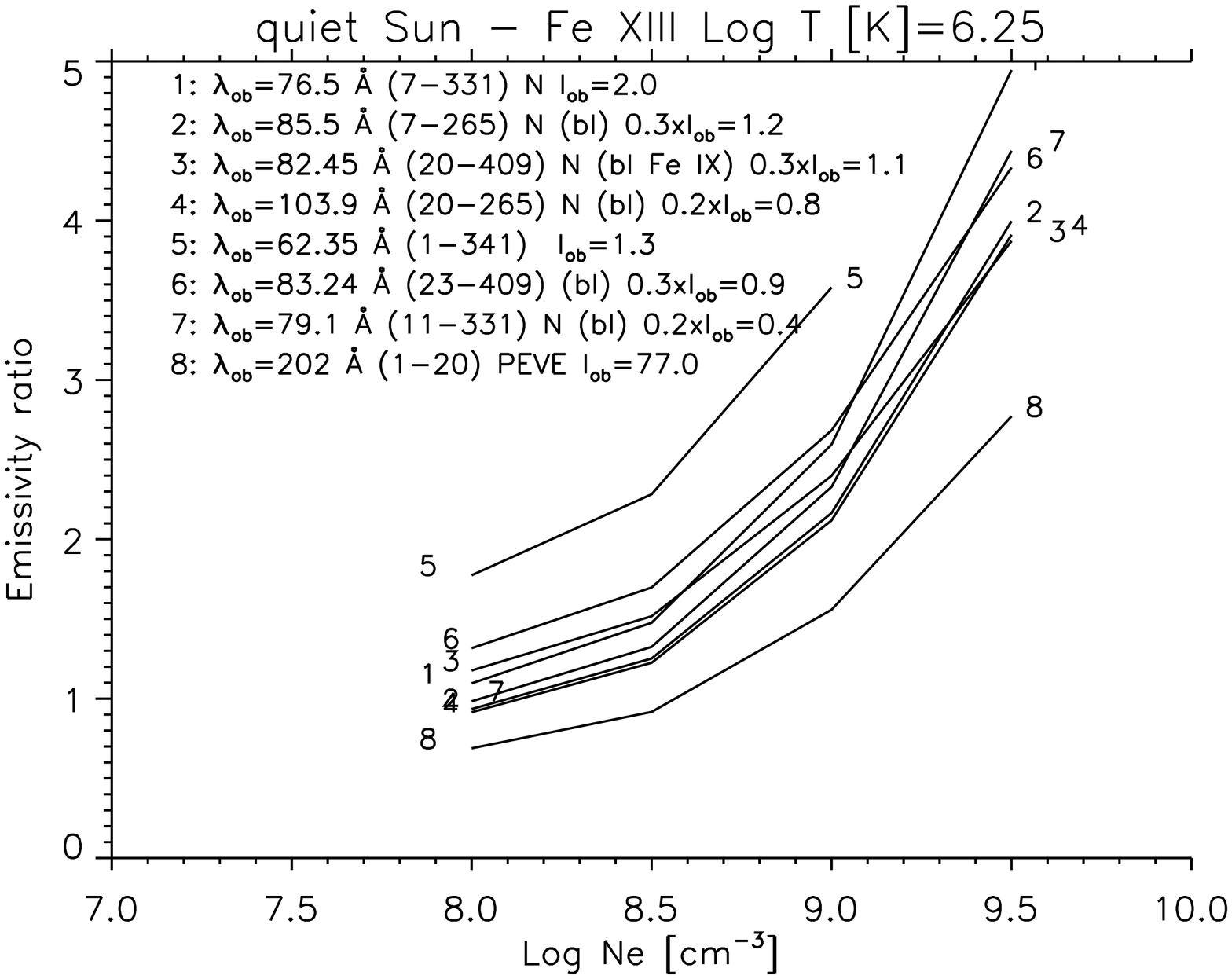, width=6.5cm,angle=90 }}
\caption{Emissivity ratio  curves  relative to  the 
main \ion{Fe}{xiii} lines. Top: A85 solar flare observation.
Bottom: M72 quiet sun observation.}
 \label{fig:fe_13}
\end{figure}

The APAP atomic data for  \ion{Fe}{xiii} have been presented in 
\cite{delzanna_storey:12_fe_13}.
Here we use the most complete atomic model, with excitation rates calculated with the 
 R-matrix  for up to the $n=4$ levels,  and DW up to $n=6$.
 Table~\ref{tab:fe_13} lists the 
 relative intensities of the brightest soft X-ray lines in \ion{Fe}{xiii}.
The previous identifications are due to Fawcett. 
\cite{kastner_etal:78} later provided some tentative identifications
of a few further lines.
\cite{vilkas_ishikawa:04_fe_13n4} presented ab-initio atomic structure calculations,
and suggested that in several cases misidentifications have 
occurred.
The energies of the lower $n=3$ levels have been carefully
assessed in \cite{delzanna:11_fe_13} and are adopted here.
 Fig.~\ref{fig:fe_13} shows the emissivity ratio  curves  relative to  the 
A85 and M72 observations.

% Fig.~\ref{fig:fe_13}

The assessment of the \ion{Fe}{xiii} soft X-ray lines has been difficult, and a more consistent
picture will need to await further observations.
Table~\ref{tab:fe_13} clearly shows that a number among the brightest lines were not
identified. Fawcett's identifications look sound in terms of wavelengths,
however a few discrepancies are present in the solar spectra. 
If one for example assumes that the 
3--344  3s$^2$ 3p$^2$ $^3$P$_{2}$--3s$^2$ 3p 4d $^3$D$_{3}$ 62.975~\AA\ identification
is correct, the model predicts for the 
1--341  3s$^2$ 3p$^2$ $^3$P$_{0}$--3s$^2$ 3p 4d $^3$D$_{1}$ 62.353~\AA\ line 
an intensity of 15 in the A85 spectrum. A85  reports a weak line at 
62.36~\AA, but does not provide an intensity. 
The 2--341  3s$^2$ 3p$^2$ $^3$P$_{1}$--3s$^2$ 3p 4d 62.717~\AA\ line 
would be blended with a stronger \ion{Mg}{ix} line.

The strongest \ion{Fe}{xiii} soft X-ray line is the 7--331  3s 3p$^3$ $^3$D$_{1}$--3s 3p$^2$ 4s $^3$P$_{0}$.
This line is strong due to a large forbidden collision strength for 
the core-excited  3s$^2$ 3p$^2$ $^3$P$_{0}$ -- 3s 3p$^2$ 4s $^3$P$_{0}$ transition.
The second decay from the upper level to the  3s 3p$^3$ $^3$P$_{1}$ (level 11) has about 1/4 the intensity
of the  the 7--331 line and ought to be observable.
The predicted wavelength for the 7--331 is around 76~\AA, where there are  three
candidate lines, at  76.113, 76.507, and 76.867~\AA\ in Be72. 
A85 wavelengths are 76.12, 76.50, 76.80~\AA, while  M72 ones are  76.14, 76.51, 76.88~\AA.
The first line was identified by A85 as a blend of \ion{Fe}{xiv} (see above)
and the \ion{Fe}{xiii} 4--221  3s$^2$ 3p$^2$ $^1$D$_{2}$--3s$^2$ 3p 4s $^1$P$_{1}$
observed by Fawcett at 76.117~\AA.
This identification is incorrect, given that the  4--221 is extremely weak 
in solar conditions. 
The A85 intensities can  account for the 76.12~\AA\ line being the  7--331, 
however in the M72 spectra the intensity of the 76.12~\AA\ line is well accounted for
by a self-blend of  \ion{Fe}{xiv} for both the active  
(as we have seen above) and quiet Sun spectrum, so this  candidate line
is discarded.

The second possibility is the 76.50~\AA\ line.
A85 identified it as an  \ion{Fe}{xvi} transition, however the 
\ion{Fe}{xvi} line only contributes 25\% of the observed intensity (see above).
It is interesting to note that the 76.50~\AA\ line is well observed also in 
quiet Sun conditions, which also confirms the fact that this line cannot be due to \ion{Fe}{xvi},
given that the quiet Sun does not have any significant  \ion{Fe}{xvi} emission.
In the A85 spectrum, there is plenty of intensity to account for 
the 7--331 line. In the M72 quiet Sun spectrum the line is  weak. 
However, good agreement is found between the M72 intensity of this line and the 
main $n=3 \to n=3$ EUV transition at 202~\AA\ observed by PEVE, as  Fig.~\ref{fig:fe_13} shows,
which is a strong argument in support of this identification.
The second decay (11--331) would fall at 79.07~\AA, where indeed a line is observed.
If this identification is correct, it means that all the other  \ion{Fe}{xiii} lines are 
severely blended.

The third possibility is the stronger (unidentified) line at 76.867~\AA.
However, in this case the second decay (11--331) would fall at 
79.46~\AA, where actually there is a line which can be 
attributed solely to \ion{Fe}{xii} (see below).
So in conclusion the 76.50~\AA\ line is favored.
There is a strong line in Fawcett's  C53 plate at 76.51~\AA. % blended with a Fe X ???

Fawcett identified a few decays from the 
3s$^2$ 3p 4p configuration, but not the two brightest ones in 
solar conditions, the decays from 3s$^2$ 3p 4p $^3$P$_{0}$ (level 265)
to 3s 3p$^3$ $^3$D$_{1}$, 3s$^2$ 3p 3d $^3$P$_{1}$ (levels 7,20).
Fawcett's energies for the  3s$^2$ 3p 4p levels suggest that 
the first decay  should be the  84.72~\AA\ line
in the M72 spectrum, however the second would fall near  103.6~\AA,
where the intensity is solely due to the  strongest \ion{Fe}{ix} soft X-ray line (see below).
A better choice are the two lines observed by Be72 at 
85.470 and 103.928~\AA. Their wavelengths are in exact agreement 
with the known energy difference between levels 7, 20, which is 
a strong argument in favor of this identification (although 
their M72 intensities are too strong).

\def\baselinestretch{1.}

\begin{table*}[!htbp]
\caption{The relative intensities of the brightest soft X-ray lines in \ion{Fe}{xii}.}
\begin{center}
%\scriptsize
\footnotesize
\begin{tabular}[c]{@{}llcrccrccccl@{}}
\hline\hline\noalign{\smallskip}  

% File: fe_13_rm4+dw6_wp_8_19_6.25_emiss
 $i$--$j$ & Levels   &  $Int$ &  $Int$ & $gf$ &  A$_{ji}$(s$^{-1}$) &   $\lambda_{\rm exp}$(\AA) & $\lambda_{\rm th}$(\AA)   & New \\   
  &  & 1.0$\times$10$^{8}$ & 1.0$\times$10$^{19}$ &  &    &  \\ 
\noalign{\smallskip}\hline\noalign{\smallskip}                                   %inserts single line

6--467 & 3s 3p$^4$ $^4$P$_{5/2}$--3s 3p$^3$ 4s $^4$S$_{3/2}$ & 1.0 & 2.3$\times$10$^{-2}$ & 0.33 & 8.0$\times$10$^{10}$ &  -  & 80.76 &  82.672  \\ 
7--467 & 3s 3p$^4$ $^4$P$_{3/2}$--3s 3p$^3$ 4s $^4$S$_{3/2}$ & 0.58 & 1.3$\times$10$^{-2}$ & 0.19 & 4.6$\times$10$^{10}$ &  -  & 81.39 &   83.336 \\
 
6--390 & 3s 3p$^4$ $^4$P$_{5/2}$--3s$^2$ 3p$^2$ 4p $^4$S$_{3/2}$ & 0.55 & 1.1$\times$10$^{-2}$ & 6.9$\times$10$^{-2}$ & 1.4$\times$10$^{10}$ &  -  & 89.03 &  91.004 \\ 

1--288 & 3s$^2$ 3p$^3$ $^4$S$_{3/2}$--3s$^2$ 3p$^2$ 4s $^4$P$_{5/2}$ & 0.55 & 3.5$\times$10$^{-2}$ & 0.29 & 5.0$\times$10$^{10}$ & 79.488 & 78.29 (-1.2) &   \\ 

7--390 & 3s 3p$^4$ $^4$P$_{3/2}$--3s$^2$ 3p$^2$ 4p $^4$S$_{3/2}$ & 0.48 & 1.0$\times$10$^{-2}$ & 6.0$\times$10$^{-2}$ & 1.2$\times$10$^{10}$ &  -  & 89.78 & 91.808  \\
 
1--278 & 3s$^2$ 3p$^3$ $^4$S$_{3/2}$--3s$^2$ 3p$^2$ 4s $^4$P$_{3/2}$ & 0.32 & 2.3$\times$10$^{-2}$ & 0.20 & 5.1$\times$10$^{10}$ & 80.022 & 78.78 (-1.2) &   \\ 

8--467 & 3s 3p$^4$ $^4$P$_{1/2}$--3s 3p$^3$ 4s $^4$S$_{3/2}$ & 0.29 & 6.7$\times$10$^{-3}$ & 9.9$\times$10$^{-2}$ & 2.3$\times$10$^{10}$ &  -  & 81.67 &  83.631 \\

8--390 & 3s 3p$^4$ $^4$P$_{1/2}$--3s$^2$ 3p$^2$ 4p $^4$S$_{3/2}$ & 0.28 & 5.8$\times$10$^{-3}$ & 3.6$\times$10$^{-2}$ & 6.9$\times$10$^{9}$ &  -  & 90.13 &  92.178 \\ 

1--484 & 3s$^2$ 3p$^3$ $^4$S$_{3/2}$--3s$^2$ 3p$^2$ 4d $^4$P$_{5/2}$ & 0.26 & 2.6$\times$10$^{-2}$ & 0.52 & 1.3$\times$10$^{11}$ & 66.297 & 65.31 (-1.0) & (bl \ion{Fe}{xvi})  \\ 
%;****

29--390 & 3s$^2$ 3p$^2$ 3d $^4$P$_{3/2}$--3s$^2$ 3p$^2$ 4p $^4$S$_{3/2}$ & 0.26 & 5.4$\times$10$^{-3}$ & 5.3$\times$10$^{-2}$ & 6.4$\times$10$^{9}$ &  -  & 114.90 & 116.76 (bl \ion{Fe}{ix}) \\ 
27--390 & 3s$^2$ 3p$^2$ 3d $^4$P$_{5/2}$--3s$^2$ 3p$^2$ 4p $^4$S$_{3/2}$ & 0.25 & 5.3$\times$10$^{-3}$ & 5.1$\times$10$^{-2}$ & 6.2$\times$10$^{9}$ &  -  & 114.37 & 116.18  \\ 

17--383 & 3s$^2$ 3p$^2$ 3d $^4$F$_{9/2}$--3s$^2$ 3p$^2$ 4p $^4$D$_{7/2}$ & 0.23 & 4.3$\times$10$^{-2}$ & 0.46 & 3.2$\times$10$^{10}$ &  108.44 & 107.04 & 109.03 ?  \\ 

1--487 & 3s$^2$ 3p$^3$ $^4$S$_{3/2}$--3s$^2$ 3p$^2$ 4d $^4$F$_{5/2}$ & 0.22 & 2.5$\times$10$^{-2}$ & 0.38 & 9.6$\times$10$^{10}$ & 66.047 & 65.10 (-0.9) &   \\ 
1--490 & 3s$^2$ 3p$^3$ $^4$S$_{3/2}$--3s$^2$ 3p$^2$ 4d $^4$P$_{3/2}$ & 0.19 & 1.7$\times$10$^{-2}$ & 0.41 & 1.5$\times$10$^{11}$ & 65.905 & 64.97 (-0.9) &   \\ 

16--370 & 3s$^2$ 3p$^2$ 3d $^4$F$_{7/2}$--3s$^2$ 3p$^2$ 4p $^4$D$_{5/2}$ & 0.16 & 3.0$\times$10$^{-2}$ & 0.34 & 3.2$\times$10$^{10}$ & 108.605 & 107.16 (-1.4) & ?  \\
 
1--272 & 3s$^2$ 3p$^3$ $^4$S$_{3/2}$--3s$^2$ 3p$^2$ 4s $^4$P$_{1/2}$ & 0.16 & 1.2$\times$10$^{-2}$ & 0.10 & 5.2$\times$10$^{10}$ & 80.515 & 79.20 (-1.3) & (bl \ion{Fe}{xiv})  \\ 

30--390 & 3s$^2$ 3p$^2$ 3d $^4$P$_{1/2}$--3s$^2$ 3p$^2$ 4p $^4$S$_{3/2}$ & 0.14 & 3.0$\times$10$^{-3}$ & 2.9$\times$10$^{-2}$ & 3.5$\times$10$^{9}$ &  -  & 115.25 &  117.2 \\ 

27--467 & 3s$^2$ 3p$^2$ 3d $^4$P$_{5/2}$--3s 3p$^3$ 4s $^4$S$_{3/2}$ & 0.14 & 3.2$\times$10$^{-3}$ & 7.1$\times$10$^{-2}$ & 1.1$\times$10$^{10}$ &  -  & 101.09 & 102.94  \\

15--364 & 3s$^2$ 3p$^2$ 3d $^4$F$_{5/2}$--3s$^2$ 3p$^2$ 4p $^4$D$_{3/2}$ & 0.13 & 1.9$\times$10$^{-2}$ & 0.23 & 3.2$\times$10$^{10}$ & 108.862 & 107.33 (-1.5) & ? 109.5 \\ 

15--619 & 3s$^2$ 3p$^2$ 3d $^4$F$_{5/2}$--3s$^2$ 3p$^2$ 4f $^4$G$_{7/2}$ & 0.13 & 5.7$\times$10$^{-2}$ & 3.78 & 4.3$\times$10$^{11}$ & 84.520 & 83.15 (-1.4) &   \\ 

1--590 & 3s$^2$ 3p$^3$ $^4$S$_{3/2}$--3s 3p$^3$ 4p $^4$P$_{5/2}$ & 0.12 & 9.8$\times$10$^{-3}$ & 0.39 & 1.1$\times$10$^{11}$ &  -  & 62.40 &   \\ 
17--644 & 3s$^2$ 3p$^2$ 3d $^4$F$_{9/2}$--3s$^2$ 3p$^2$ 4f $^4$G$_{11/2}$ & 0.10 & 9.9$\times$10$^{-2}$ & 6.57 & 5.1$\times$10$^{11}$ & 84.520 & 83.24 (-1.3) &   \\ 
2--491 & 3s$^2$ 3p$^3$ $^2$D$_{3/2}$--3s$^2$ 3p$^2$ 4d $^2$F$_{5/2}$ & 9.3$\times$10$^{-2}$ & 2.9$\times$10$^{-2}$ & 0.51 & 1.2$\times$10$^{11}$ & 67.821 & 66.81 (-1.0) &   \\

\noalign{\smallskip}
1--30 & 3s$^2$ 3p$^3$ $^4$S$_{3/2}$--3s$^2$ 3p$^2$ 3d $^4$P$_{1/2}$ & 21. & 0.20 & 1.00 & 8.8$\times$10$^{10}$ & 192.394 & 188.87 (-3.5) &   \\ 

\noalign{\smallskip}\hline\noalign{\smallskip}                                   %inserts single line
\end{tabular}
\normalsize
\tablefoot{The relative line intensities  (photons)  $Int=N_{j} A_{ji}/N_{\rm e}$ 
were calculated at log N$_{\rm e}$ [cm$^{-3}$]=8,19 and log $T$$_{\rm e}$ [K]= 6.2
}
\end{center}
\label{tab:fe_12}
\end{table*}
% Table~\ref{tab:fe_12}

%\def\baselinestretch{1.5}

Fawcett also identified a few decays from the 
3s$^2$ 3p 4f configuration, but not the brightest ones for the solar corona,
the 20--409 and 23--409 lines.  They are identified here as the 
lines at 82.45 and 83.24~\AA\ in the M72 spectrum.
The first is blended with an \ion{Fe}{ix} line, the second 
is also blended. Be72 provides  the 82.425 and 83.221~\AA\ wavelengths,
in excellent agreement with the known energies of levels 20, 23
(providing  1708166 and 1708125 cm$^{-1}$ as energies for the 3s$^2$ 3p 4f $^3$F$_{2}$).

\subsection{\ion{Fe}{xii}}

\begin{figure*}[!htbp]
\sidecaption 
\centerline{\epsfig{file=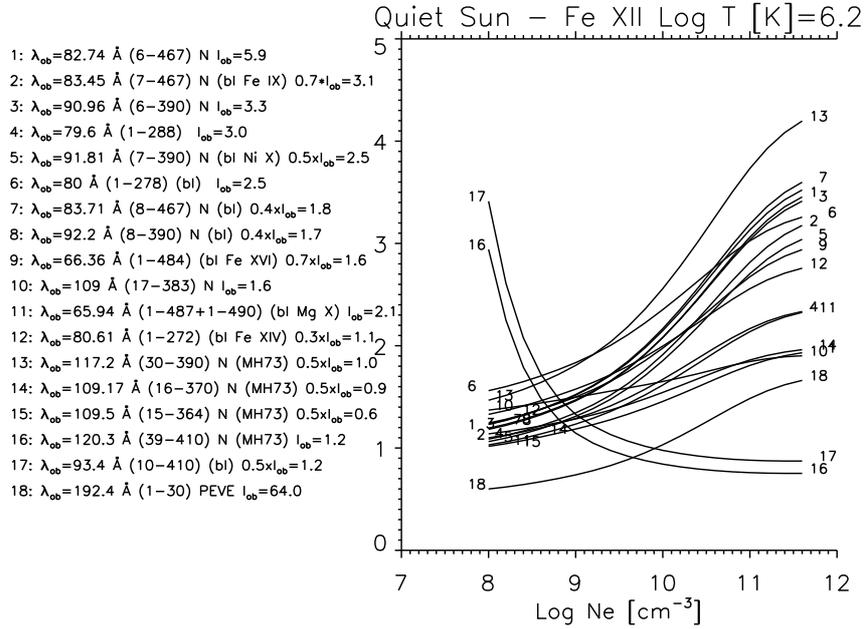, width=8.5cm,angle=90 }}
\caption{Emissivity ratio  curves  relative to  the 
main \ion{Fe}{xii} lines and the  M72 quiet Sun observation.}
 \label{fig:fe_12}
\end{figure*}

The APAP atomic data for  \ion{Fe}{xii} have been presented in 
\cite{delzanna_etal:12_fe_12}.
We use the most complete atomic model, with excitation rates calculated with the 
 R-matrix method  for up to the $n=4$ levels,  and DW up to $n=6$.
 Table~\ref{tab:fe_12} lists the 
 relative intensities of the brightest soft X-ray lines in \ion{Fe}{xii}.
Previous identifications are due to Fawcett.
The energies of the lower $n=3$ levels have been carefully
assessed in \cite{delzanna_mason:05_fe_12} and are adopted here.
 Fig.~\ref{fig:fe_12} shows the emissivity ratio  curves  relative to  the 
 M72 observations.

% Fig.~\ref{fig:fe_12}

As we have seen for the other ions, the core-excited 
3s$^2$ 3p$^3$ $^4$S$_{3/2}$--3s 3p$^3$ 4s $^4$S$_{3/2}$ is a strong 
forbidden transition which provides a large population to the upper level (467),which 
in turn decays to levels 6,7,8,27,29. The first three decays are strong, indeed
as shown in Table~\ref{tab:fe_12} the first two are the strongest soft X-ray lines
from this ion.

\def\baselinestretch{1.}

\begin{table*}[!htbp]
\caption{The relative intensities of the brightest soft X-ray lines in \ion{Fe}{xi}.}
\begin{center}
%\scriptsize
\footnotesize
\begin{tabular}[c]{@{}llcrccrccccl@{}}
\hline\hline\noalign{\smallskip}  

% File: fe_13_rm4+dw6_wp_8_19_6.25_emiss
 $i$--$j$ & Levels   &  $Int$ &  $Int$ & $gf$ &  A$_{ji}$(s$^{-1}$) &   $\lambda_{\rm exp}$(\AA) & $\lambda_{\rm th}$(\AA)   & New \\   
  &  & 1.0$\times$10$^{8}$ & 1.0$\times$10$^{20}$ &  &    &  \\ 
\noalign{\smallskip}\hline\noalign{\smallskip}                                   %inserts single line

6--596 & 3s 3p$^5$ $^3$P$_{2}$--3s 3p$^4$ 4s $^3$P$_{2}$ & 1.0 & 4.3$\times$10$^{-3}$ & 0.46 & 7.7$\times$10$^{10}$ &  -  & 85.51 &  88.933 \\ 
6--454 & 3s 3p$^5$ $^3$P$_{2}$--3s$^2$ 3p$^3$ 4p $^3$P$_{2}$ & 0.75 & 1.3$\times$10$^{-3}$ & 8.3$\times$10$^{-2}$ & 1.1$\times$10$^{10}$ &  -  & 96.36 &  100.575  \\ 
1--291 & 3s$^2$ 3p$^4$ $^3$P$_{2}$--3s$^2$ 3p$^3$ 4s $^3$D$_{3}$ & 0.53 & 7.3$\times$10$^{-3}$ & 0.37 & 4.5$\times$10$^{10}$ & 86.772 & 84.52 (-2.3) &  86.765 \\ 
1--265 & 3s$^2$ 3p$^4$ $^3$P$_{2}$--3s$^2$ 3p$^3$ 4s $^3$S$_{1}$ & 0.32 & 4.9$\times$10$^{-3}$ & 0.39 & 1.1$\times$10$^{11}$ & 89.185 & 86.85 (-2.3) &  89.178 \\ 
42--454 & 3s$^2$ 3p$^3$ 3d $^3$D$_{3}$--3s$^2$ 3p$^3$ 4p $^3$P$_{2}$ & 0.28 & 4.8$\times$10$^{-4}$ & 5.8$\times$10$^{-2}$ & 4.1$\times$10$^{9}$ &  -  & 132.77 & 138.21  \\ 
1--536 & 3s$^2$ 3p$^4$ $^3$P$_{2}$--3s$^2$ 3p$^3$ 4d $^3$D$_{3}$ & 0.28 & 8.8$\times$10$^{-3}$ & 0.73 & 1.3$\times$10$^{11}$ & 72.635 & 70.98 (-1.7) &   \\ 
7--596 & 3s 3p$^5$ $^3$P$_{1}$--3s 3p$^4$ 4s $^3$P$_{2}$ & 0.25 & 1.1$\times$10$^{-3}$ & 0.12 & 1.9$\times$10$^{10}$ &  -  & 86.20 &  89.703 \\
 
7--454 & 3s 3p$^5$ $^3$P$_{1}$--3s$^2$ 3p$^3$ 4p $^3$P$_{2}$ & 0.23 & 3.9$\times$10$^{-4}$ & 2.6$\times$10$^{-2}$ & 3.3$\times$10$^{9}$ &  -  & 97.25 &  101.559 \\ 

16--749 & 3s$^2$ 3p$^3$ 3d $^3$D$_{3}$--3s$^2$ 3p$^3$ 4f $^3$F$_{4}$ & 0.21 & 1.6$\times$10$^{-2}$ & 3.40 & 2.9$\times$10$^{11}$ &  -  & 89.77 & ? 92.18 \\ 

38--454 & 3s$^2$ 3p$^3$ 3d $^3$P$_{2}$--3s$^2$ 3p$^3$ 4p $^3$P$_{2}$ & 0.21 & 3.6$\times$10$^{-4}$ & 4.1$\times$10$^{-2}$ & 3.0$\times$10$^{9}$ &  -  & 128.40 & 133.95  \\ 
42--813 & 3s$^2$ 3p$^3$ 3d $^3$D$_{3}$--3s$^2$ 3p$^3$ 4f $^3$F$_{4}$ & 0.17 & 8.7$\times$10$^{-3}$ & 2.31 & 1.6$\times$10$^{11}$ &  -  & 99.64 & ? 102.10  \\ 

16--377 & 3s$^2$ 3p$^3$ 3d $^3$D$_{3}$--3s$^2$ 3p$^3$ 4p $^3$P$_{2}$ & 0.16 & 1.6$\times$10$^{-3}$ & 0.16 & 1.3$\times$10$^{10}$ &  -  & 119.93 & ? 124.72  \\ 
24--424 & 3s$^2$ 3p$^3$ 3d $^3$G$_{5}$--3s$^2$ 3p$^3$ 4p $^3$F$_{4}$ & 0.15 & 4.9$\times$10$^{-3}$ & 0.47 & 2.2$\times$10$^{10}$ & 123.490 & 120.03 (-3.5) & ? 124.72 \\ 

2--265 & 3s$^2$ 3p$^4$ $^3$P$_{1}$--3s$^2$ 3p$^3$ 4s $^3$S$_{1}$ & 0.15 & 2.2$\times$10$^{-3}$ & 0.18 & 4.8$\times$10$^{10}$ & 90.204 & 87.79 (-2.4) & 90.17  \\ 
1--289 & 3s$^2$ 3p$^4$ $^3$P$_{2}$--3s$^2$ 3p$^3$ 4s $^3$D$_{2}$ & 0.14 & 3.2$\times$10$^{-3}$ & 0.16 & 2.8$\times$10$^{10}$ & 87.025 & 84.71 (-2.3) & 87.018  \\
 
39--454 & 3s$^2$ 3p$^3$ 3d $^3$S$_{1}$--3s$^2$ 3p$^3$ 4p $^3$P$_{2}$ & 0.12 & 2.1$\times$10$^{-4}$ & 2.4$\times$10$^{-2}$ & 1.8$\times$10$^{9}$ &  -  & 129.08 &  134.34 \\ 
4--295 & 3s$^2$ 3p$^4$ $^1$D$_{2}$--3s$^2$ 3p$^3$ 4s $^1$D$_{2}$ & 0.12 & 9.3$\times$10$^{-3}$ & 0.71 & 1.2$\times$10$^{11}$ & 89.044 & 86.68 (-2.4) &  89.087  \\ 

30--460 & 3s$^2$ 3p$^3$ 3d $^3$F$_{3}$--3s$^2$ 3p$^3$ 4p $^3$D$_{2}$ & 0.12 & 1.9$\times$10$^{-3}$ & 0.20 & 1.7$\times$10$^{10}$ & 123.572 & 120.21 (-3.4) & ? 125.40  \\ 

14--353 & 3s$^2$ 3p$^3$ 3d $^5$D$_{4}$--3s$^2$ 3p$^3$ 4p $^5$P$_{3}$ & 0.12 & 3.8$\times$10$^{-3}$ & 0.39 & 2.4$\times$10$^{10}$ & 121.419 & 118.00 (-3.4) & ? 123.49  \\

14--704 & 3s$^2$ 3p$^3$ 3d $^5$D$_{4}$--3s$^2$ 3p$^3$ 4f $^5$F$_{5}$ & 9.3$\times$10$^{-2}$ & 2.8$\times$10$^{-2}$ & 6.23 & 4.3$\times$10$^{11}$ & 91.733 & 89.18 (-2.6) &   \\

 \noalign{\smallskip}
1--42 & 3s$^2$ 3p$^4$ $^3$P$_{2}$--3s$^2$ 3p$^3$ 3d $^3$D$_{3}$ & 55. & 4.9$\times$10$^{-2}$ & 4.82 & 1.4$\times$10$^{11}$ & 180.401 & 176.36 (-4.0) &   \\ 

\noalign{\smallskip}\hline\noalign{\smallskip}                                   %inserts single line
\end{tabular}
\normalsize
\tablefoot{The relative line intensities  (photons)  $Int=N_{j} A_{ji}/N_{\rm e}$ 
were calculated at log N$_{\rm e}$ [cm$^{-3}$]=8, 20 and log $T$$_{\rm e}$ [K]= 6.15
}
\end{center}
\label{tab:fe_11}
\end{table*}
% Table~\ref{tab:fe_11}

%\def\baselinestretch{1.5}

Some decays from the 3s$^2$ 3p$^2$ 4s were identified by Fawcett. 
The differences between observed and predicted wavelengths are about 1.2~\AA.
If similar differences were applied to the 3s 3p$^3$ 4s configuration, we
obtain a predicted wavelength for the strongest 6--467 transition of 82~\AA.
There are a few candidate lines both in solar and laboratory spectra, however
the strongest one is observed by M72 at 82.75~\AA.
Fawcett's  plate C53 also has a strong broad line around 82.74~\AA.
Be72 provided a wavelength of 82.672~\AA\ for the same line. 
Based on this, the second and third decays (7,8 --467)
are predicted to be at 83.336 and 83.635~\AA, in excellent agreement with the lines
observed by Be72 at 83.336 and 83.631~\AA. This is an unlikely coincidence and confirms 
the present identification. 
The 83.336~\AA\ at the M72 resolution is blended with an \ion{Fe}{ix} transition
(see below). In Fawcett plate C53 there is a pseudo-continuum of transitions 
between 83.3 and 83.7~\AA, where these two decays are.

The next strongest transitions are the decays from the 3s$^2$ 3p$^2$ 4p $^4$S$_{3/2}$ (level 390)
to levels 6, 7,8, 29, 27. The level is relatively pure (78\%). 
Fawcett identified a few 3s$^2$ 3p$^2$ 4p levels, and the difference between predicted and 
observed wavelengths is around 1.4~\AA. The strongest decay (6--390) should then fall 
around 90.4~\AA. There is a weak line at 90.4~\AA\ in M72 (90.503~\AA\ in Be72),
but the corresponding decays to levels 7,8 would then be 
at  91.30 and 91.66~\AA\ (using Be72 wavelength). There is no line at 
91.30~\AA.
Of all the lines around 90.4 there is only one with the appropriate wavelength,
observed by Be72 at 91.004~\AA. This wavelength predicts decays to levels 7,8 
at  91.809 and   92.172~\AA. Be72 observed two lines at 91.808 and   92.178~\AA,
an unlikely coincidence, although both lines would have to 
be blended, the first one with \ion{Ni}{x} and \ion{Fe}{xi}. 
The difference between observed and predicted wavelength  with the new identifications is 2~\AA.
The weaker decays to levels 29,27 would fall at 116.76 and  116.18~\AA,
and would be blended with other stronger transitions.

One question then naturally arises: are the other identifications of the 
3s$^2$ 3p$^2$ 4p levels correct ?
The solar spectra cannot resolve this issue. The best solar spectrum at the
wavelengths of these decays is the MH73, but the spectral resolution is not enough.
Be72 does not list the (weak) lines observed by MH73.
The Fawcett plate does provide viable alternatives for all the main 
transitions, with observed wavelengths about 2~\AA\ away from the predicted ones,
so it is possible that all previous identifications are incorrect.

The M72 intensities are in excellent agreement with predictions and the 
present identifications, for the three strongest decays from the 
3s 3p$^3$ 4s $^4$S$_{3/2}$ (6-467), 3s$^2$ 3p$^2$ 4p $^4$S$_{3/2}$ (6--390), and 
3p$^2$ 4s $^4$P$_{5/2}$ (1-288), as Fig.~\ref{fig:fe_12} shows.
The intensity of the self-blend (at the M72 resolution) of the 
3s$^2$ 3p$^2$ 4d $^4$F$_{5/2}$,$^4$P$_{3/2}$ decays to the ground state
is also in excellent agreement.
There is a discrepancy with the EUV lines as measured with PEVE, however
(cf. the 192.4~\AA\ in  Fig.~\ref{fig:fe_12}).
This discrepancy could in part be due to the lower solar activity
during 2008, but also in part due to residual blending in all the \ion{Fe}{xii} lines.

Finally, a few remarks about some weaker lines.
The 80.50~\AA\ strong line in Fawcett C53 spectrum was identified 
as a self-blend of \ion{Fe}{xii} lines, however as 
we have seen above we predict a new strong \ion{Fe}{xiv} transition
at this wavelength. The M72 intensity supports this, given that 
the 1--272 transition is predicted to contribute only 
about 30\% to the intensity of the observed line in the solar
spectrum (60\% was estimated to be due to \ion{Fe}{xiv}).

The model predicts two weak decays from 
the 3s$^2$ 3p$^2$ 4p $^2$D$_{5/2}$ (410) level. The energy splitting
for nearby  3s$^2$ 3p$^2$ 4p levels suggests that the two decays 
should be the lines at 120.3, 93.4~\AA, observed in the MH73 and 
Fawcett's C53 plate (a possible alternative would be 121.1, 93.8~\AA).
M72 also observed a line at 93.46~\AA, listed as blended.
The MH73 intensity has approximately the right intensity.
We can then estimate the intensity of the 10-410 line 
to be about 1$\times 10^6$ phot cm$^{-2}$ s$^{-1}$, i.e. about
half of the M72 intensity.

\subsection{\ion{Fe}{xi}}

\begin{figure}[!htbp]
\centerline{\epsfig{file=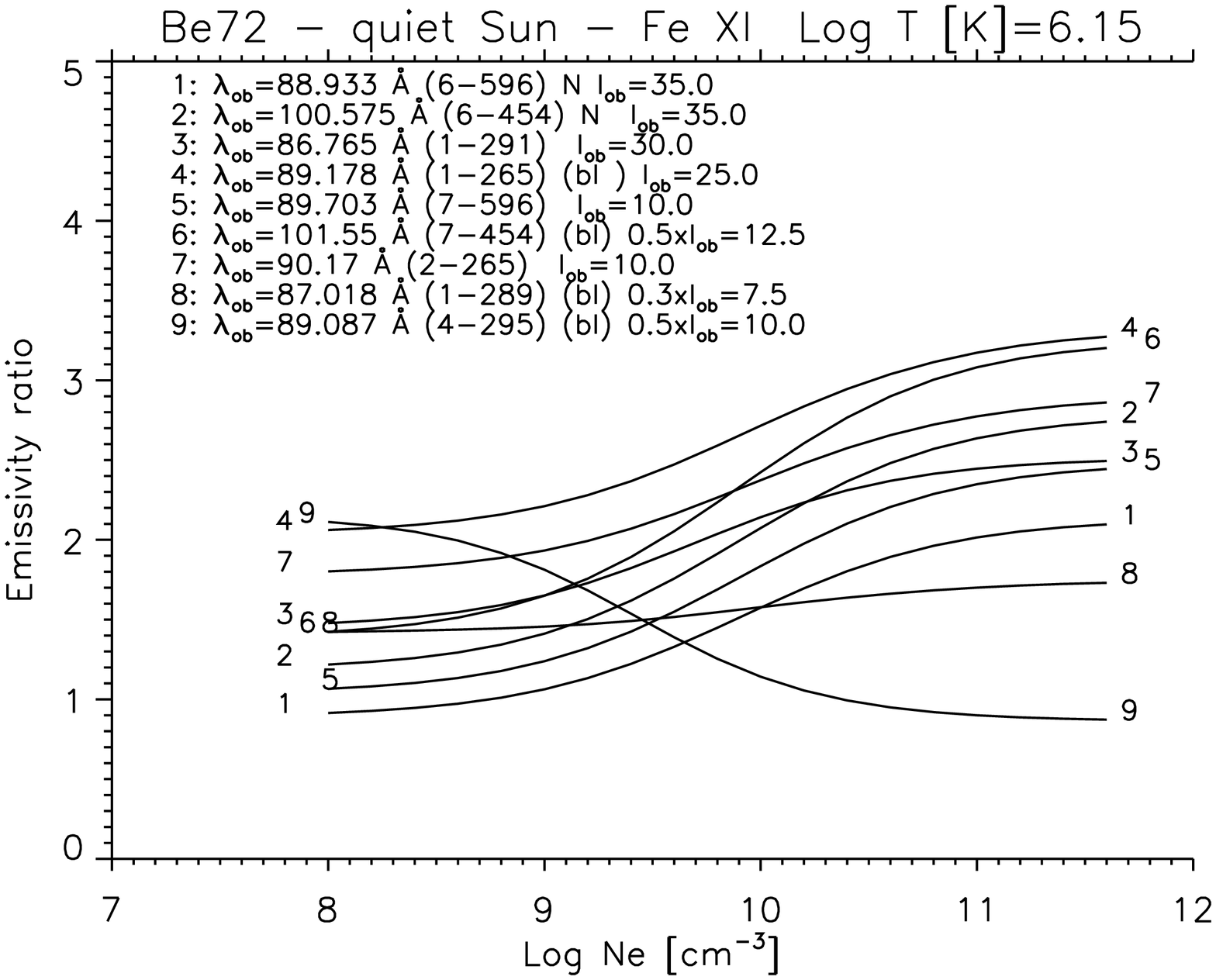, width=6.5cm,angle=90 }}
\centerline{\epsfig{file=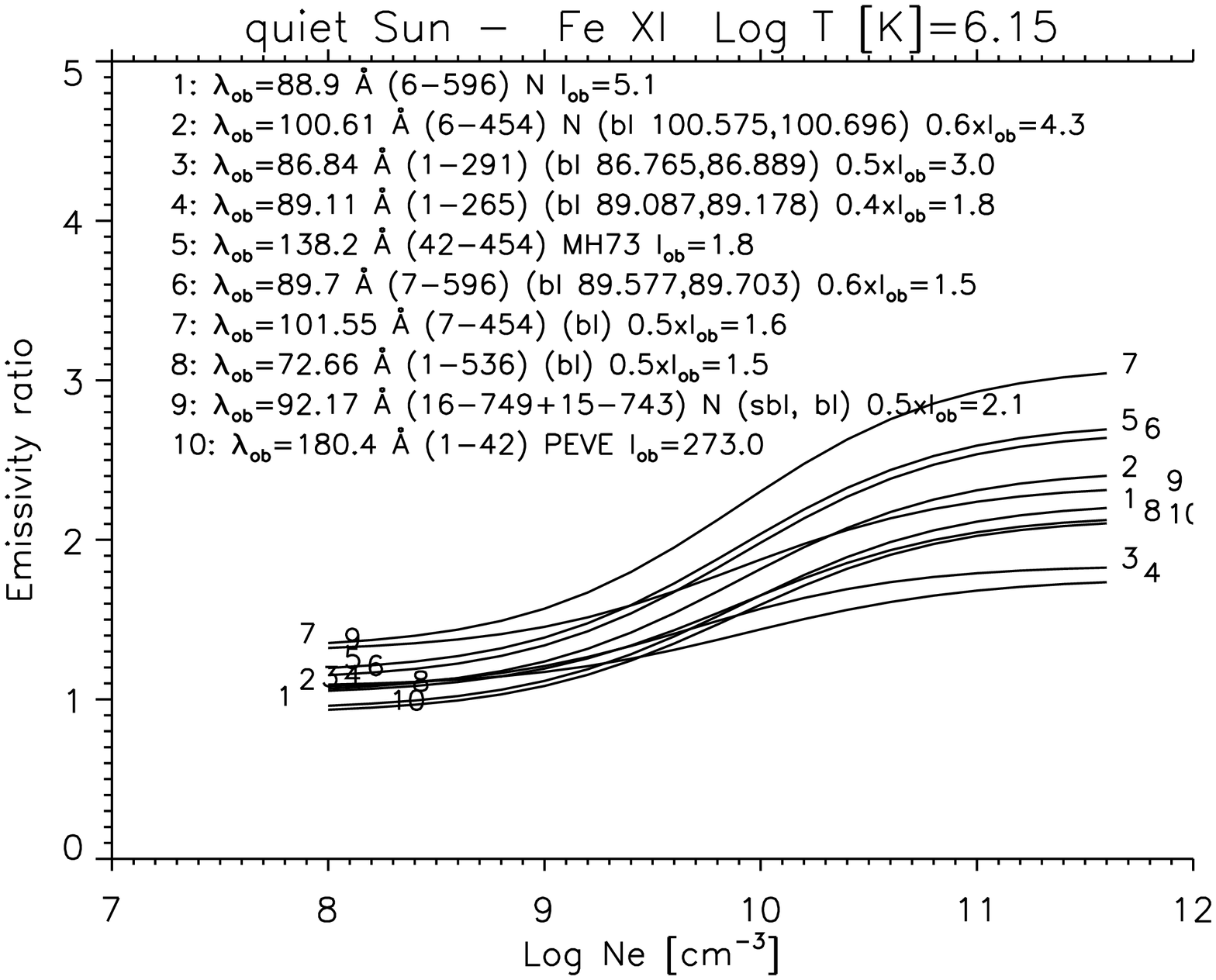, width=6.5cm,angle=90 }}
\caption{Emissivity ratio  curves  relative to  the 
main \ion{Fe}{xi} lines. Top: Be72 quiet Sun observation.
Bottom:  quiet Sun irradiances from M72, MH73 and PEVE.}
 \label{fig:fe_11}
\end{figure}
% Fig.~\ref{fig:fe_11}

The APAP atomic data for  \ion{Fe}{xi} have been presented in 
\cite{delzanna_storey:12_fe_11}.
We use the most complete atomic model, with excitation rates calculated with the 
 R-matrix method  for up to the $n=4$ levels,  and DW up to $n=6$.
 Table~\ref{tab:fe_11} lists the 
 relative intensities of the brightest soft X-ray lines in \ion{Fe}{xi}.
The previous identifications are due to \cite{edlen:37_s-like}
and Fawcett.
The energies of the lower $n=3$ levels have been carefully
assessed in \cite{delzanna:10_fe_11} and are adopted here.

 Fig.~\ref{fig:fe_11} shows the emissivity ratio  curves  relative to  the 
quiet Sun  observations.
As in the  \ion{Fe}{xiv},  \ion{Fe}{xiii}, and  \ion{Fe}{xii} case,
the strongest soft X-ray line is the unidentified dipole-allowed decay (6--596)
from a level (3s 3p$^4$ 4s $^3$P$_{2}$) that is core-excited via a strong forbidden transition
form the ground state.
On the basis of the predicted vs. observed  wavelengths of the few decays 
from the 3s$^2$ 3p$^3$ 4s  identified by Fawcett, we expect the 
6--596 line to fall around 88~\AA. There is indeed a strong line at 
88.082~\AA\ in Be72 (88.1~\AA\ in M72), however this is a \ion{Ne}{viii} transition.
Furthermore, the next decay from level 596 is the 7--596, predicted to be 
about 1/4 the intensity of the strongest line. The 88.082~\AA\ wavelength
would predict the  7--596 to fall at 88.834~\AA\ where no line is observed.
The next strongest line is the unidentified 88.933~\AA\ in Be72,
which predicts a wavelength for the 7--596 line of 89.699~\AA.
Indeed in Be72 there is a line of the right intensity (see  Fig.~\ref{fig:fe_11} top) at 89.703~\AA,
which would be a very  unlikely coincidence.
The 6--596 transition is therefore identified with the 
88.933~\AA\ line.

\def\baselinestretch{1.}

\begin{table*}[!htbp]
\caption{The relative intensities of the brightest soft X-ray lines in \ion{Fe}{x}.}
\begin{center}
%\scriptsize
\footnotesize
\begin{tabular}[c]{@{}llcrccrccccl@{}}
\hline\hline\noalign{\smallskip}  

% File: fe_13_rm4+dw6_wp_8_19_6.25_emiss
 $i$--$j$ & Levels   &  $Int$ &  $Int$ & $gf$ &  A$_{ji}$(s$^{-1}$) &   $\lambda_{\rm exp}$(\AA) & $\lambda_{\rm th}$(\AA)   & New \\   
  &  & 1.0$\times$10$^{8}$ & 1.0$\times$10$^{20}$ &  &    &  \\ 
\noalign{\smallskip}\hline\noalign{\smallskip}                                   %inserts single line

3--429 & 3s 3p$^6$ $^2$S$_{1/2}$--3s 3p$^5$ 4s $^2$P$_{3/2}$ & 1.0 & 3.2$\times$10$^{-3}$ & 0.27 & 4.8$\times$10$^{10}$ &  -  & 91.48 &  96.007 \\ 

1--202 & 3s$^2$ 3p$^5$ $^2$P$_{3/2}$--3s$^2$ 3p$^4$ 4s $^2$D$_{5/2}$ & 0.71 & 6.0$\times$10$^{-3}$ & 0.31 & 3.7$\times$10$^{10}$ & 94.012 & 90.46 (-3.5) &   \\ 
1--183 & 3s$^2$ 3p$^5$ $^2$P$_{3/2}$--3s$^2$ 3p$^4$ 4s $^2$P$_{3/2}$ & 0.63 & 9.7$\times$10$^{-3}$ & 0.50 & 8.7$\times$10$^{10}$ & 96.121 & 92.43 (-3.7) &   \\ 
1--174 & 3s$^2$ 3p$^5$ $^2$P$_{3/2}$--3s$^2$ 3p$^4$ 4s $^4$P$_{5/2}$ & 0.36 & 2.1$\times$10$^{-4}$ & 1.1$\times$10$^{-2}$ & 1.2$\times$10$^{9}$ & 97.838 & 94.20 (-3.6) &   \\ 
1--179 & 3s$^2$ 3p$^5$ $^2$P$_{3/2}$--3s$^2$ 3p$^4$ 4s $^4$P$_{3/2}$ & 0.30 & 1.7$\times$10$^{-3}$ & 8.6$\times$10$^{-2}$ & 1.4$\times$10$^{10}$ & 97.122 & 93.53 (-3.6) &   \\
 
27--302 & 3s$^2$ 3p$^4$ 3d $^2$S$_{1/2}$--3s$^2$ 3p$^4$ 4p $^2$P$_{3/2}$ & 0.25 & 4.9$\times$10$^{-4}$ & 8.2$\times$10$^{-2}$ & 5.8$\times$10$^{9}$ &  -  & 146.56 & ? 151.42  \\ 

22--267 & 3s$^2$ 3p$^4$ 3d $^2$G$_{9/2}$--3s$^2$ 3p$^4$ 4p $^2$F$_{7/2}$ & 0.20 & 2.5$\times$10$^{-3}$ & 0.35 & 1.4$\times$10$^{10}$ & 139.869 & 135.95 (-3.9) &   \\ 

2--203 & 3s$^2$ 3p$^5$ $^2$P$_{1/2}$--3s$^2$ 3p$^4$ 4s $^2$D$_{3/2}$ & 0.20 & 4.8$\times$10$^{-3}$ & 0.26 & 4.5$\times$10$^{10}$ & 95.374 & 91.70 (-3.7) &   \\ 

3--302 & 3s 3p$^6$ $^2$S$_{1/2}$--3s$^2$ 3p$^4$ 4p $^2$P$_{3/2}$ & 0.19 & 3.7$\times$10$^{-4}$ & 3.3$\times$10$^{-2}$ & 4.4$\times$10$^{9}$ &  -  & 104.65 & ? 109.52 \\

28--508 & 3s$^2$ 3p$^4$ 3d $^2$P$_{3/2}$--3s$^2$ 3p$^4$ 4f $^2$D$_{5/2}$ & 0.15 & 1.3$\times$10$^{-2}$ & 1.92 & 1.6$\times$10$^{11}$ &  -  & 112.54 & ? 113.8  \\ 

8--243 & 3s$^2$ 3p$^4$ 3d $^4$F$_{9/2}$--3s$^2$ 3p$^4$ 4p $^4$D$_{7/2}$ & 0.15 & 2.6$\times$10$^{-3}$ & 0.33 & 1.4$\times$10$^{10}$ & 140.296 & 136.05 (-4.2) &   \\ 
2--192 & 3s$^2$ 3p$^5$ $^2$P$_{1/2}$--3s$^2$ 3p$^4$ 4s $^2$P$_{1/2}$ & 0.14 & 3.5$\times$10$^{-3}$ & 0.19 & 6.4$\times$10$^{10}$ & 96.786 & 93.00 (-3.8) &   \\

1--192 & 3s$^2$ 3p$^5$ $^2$P$_{3/2}$--3s$^2$ 3p$^4$ 4s $^2$P$_{1/2}$ & 0.10 & 2.6$\times$10$^{-3}$ & 0.13 & 4.7$\times$10$^{10}$ & 95.339 & 91.68 (-3.7) &   \\

\noalign{\smallskip}
1--30 & 3s$^2$ 3p$^5$ $^2$P$_{3/2}$--3s$^2$ 3p$^4$ 3d $^2$D$_{5/2}$ & 63. & 7.9$\times$10$^{-2}$ & 5.44 & 1.9$\times$10$^{11}$ & 174.531 & 163.29 (-11.2) &   \\

\noalign{\smallskip}\hline\noalign{\smallskip}                                   %inserts single line
\end{tabular}
\normalsize
\tablefoot{The relative line intensities  (photons)  $Int=N_{j} A_{ji}/N_{\rm e}$ 
were calculated at log N$_{\rm e}$ [cm$^{-3}$]=8,20 and log $T$$_{\rm e}$ [K]= 6.0
}
\end{center}
\label{tab:fe_10}
\end{table*}
% Table~\ref{tab:fe_10}

%\def\baselinestretch{1.5}

The strongest decays from the 3s$^2$ 3p$^3$ 4s  identified by Fawcett are the 
86.765, 89.178~\AA\ lines. These and others among the strongest 
transitions are severely blended in M72, but fall within a few \AA, so it 
is reasonable to use the Be72 approximate counts to check the relative
intensities of these lines.  Fig.~\ref{fig:fe_11} (top) shows that the intensity of the 
86.772~\AA\ is in good agreement with that of the 6--596 as we have identified it,
while the 89.185~\AA\ line would be blended. Other lines such as the 
2--265 at 90.17~\AA, 1--289 at 87.018~\AA, 4-295 at 89.087~\AA\ also 
appear to be blended.

Some among the brightest lines for this ion are from the 3s$^2$ 3p$^3$ 4p,
in particular the 6--454, the second strongest transition.
Fawcett identified a few transitions in the C53 plate, 
but not the strongest ones in solar conditions. Fawcett's identifications
suggest that the 6--454 transition should fall around 99.4~\AA,
however there are no strong lines there. The strongest nearby line 
is the previously unidentified 100.575~\AA\ one.
Its Be72 intensity is in remarkable agreement with the predicted one.
The upper level has a series of decays, the main ones to levels 
42, and 7, with predicted wavelengths of 138.215 and 101.556~\AA. 
Be72 has indeed a line at 101.559~\AA\ (probably blended), 
and the MH73 spectrum has a line at 138.2~\AA\ with the right
intensity (see  Fig.~\ref{fig:fe_11}).
The new energy for level 454 (3s$^2$ 3p$^3$ 4p $^3$P$_{2}$) is significantly 
(by 11000 cm$^{-1}$) at odds with those of the levels 
identified by Fawcett (in terms of energy difference between 
observed and predicted). For each of the lines identified by Fawcett,
there are alternative candidates in the same C53 plate which 
have similar energy differences as the 3s$^2$ 3p$^3$ 4p $^3$P$_{2}$.
Some of these alternative identifications are listed in  Table~\ref{tab:fe_11}. 

Fawcett identified a few amongst the decays from the 3s$^2$ 3p$^3$ 4f,
some only tentatively. The two strongest lines in solar conditions 
were not identified. Based on the energy differences of the identified ones,
the two 16--749  and 42--813 transitions are tentatively identified with the 
92.18 and 102.10~\AA\ lines, observed in the solar spectra.

The Be72 spectrum helps in accounting for the various blends in the 
lower-resolution M72 spectrum, and good agreement is also found there,
as Fig.~\ref{fig:fe_11} (bottom) shows.
Moreover, the quiet-Sun PEVE intensity of the strongest EUV line
is in excellent agreement aswell, further confirming the atomic calculations
and the identifications. This is remarkable.

\subsection{\ion{Fe}{x}}

\begin{figure}[!htbp]
\centerline{\epsfig{file=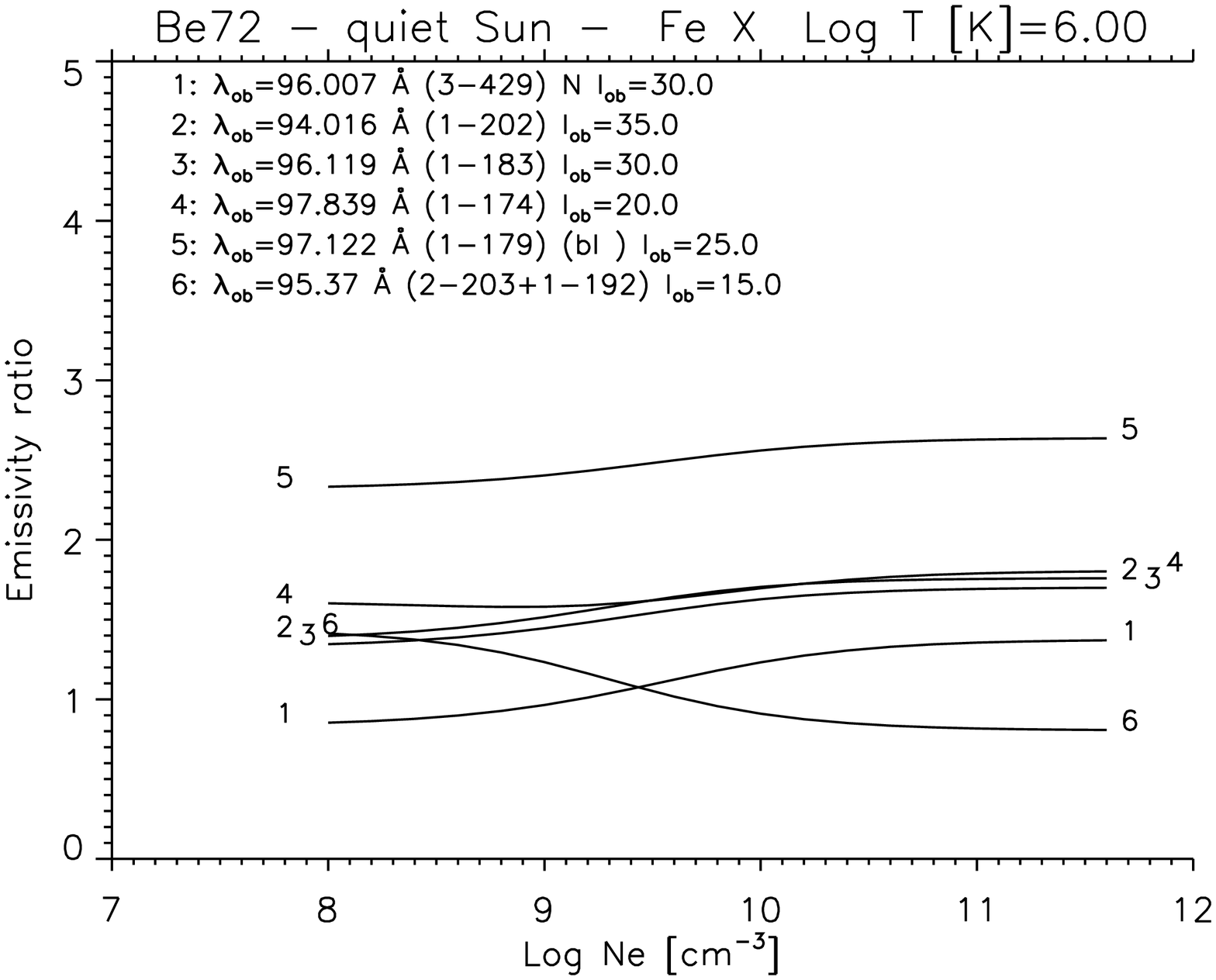, width=6.5cm,angle=90 }}
\centerline{\epsfig{file=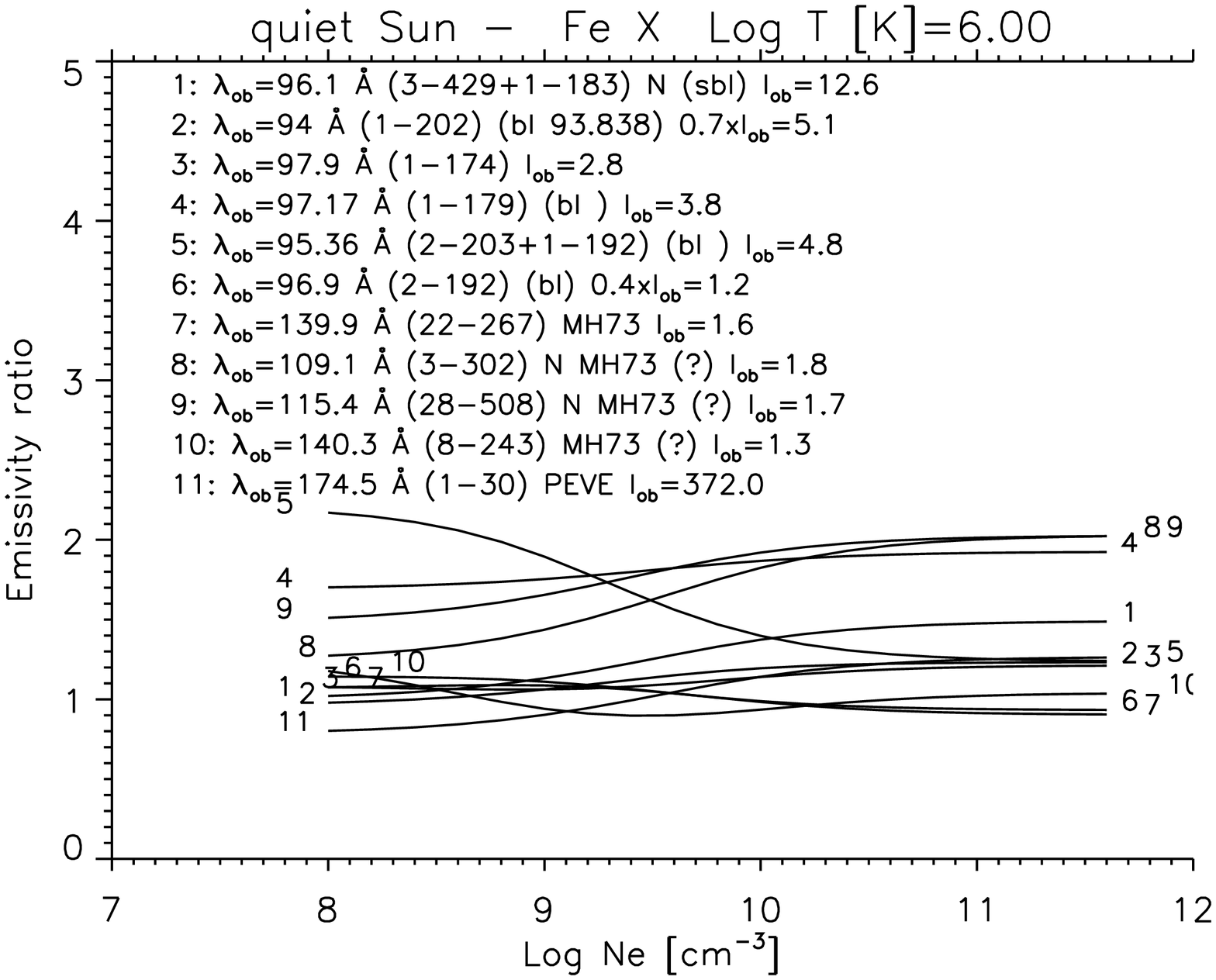, width=6.5cm,angle=90 }}
\caption{Emissivity ratio  curves  relative to  the 
main \ion{Fe}{x} lines in the quiet Sun. Top: Be72. 
Bottom: from the M72, MH73 and PEVE irradiances.}
 \label{fig:fe_10}
\end{figure}
% Fig.~\ref{fig:fe_10}

The atomic data for  \ion{Fe}{x} have been presented and discussed
in detail in \cite{delzanna_etal:12_fe_10}. We use the most 
complete atomic model, with excitation rates calculated with the 
 R-matrix  for up to $n=4$ and DW up to $n=6$.
 Table~\ref{tab:fe_10} lists the 
 relative intensities of the brightest soft X-ray lines in \ion{Fe}{x}.
The previous identifications are due to \cite{edlen:37_cl-like}
and Fawcett.
The energies of the lower $n=3$ levels have been carefully
assessed in Paper~I  and are adopted here.

 Fig.~\ref{fig:fe_10} shows the emissivity ratio  curves  relative to  the 
Be72 and M72 observations. For some of the weaker lines, the intensities
obtained from the MH73 spectrum are used.
There is good overall agreement among the main decays from the 
3s$^2$ 3p$^4$ 4s (identified by \citealt{edlen:37b}), with the exception of the 97.122~\AA\ line
which appears to be blended,  even at the Be72 resolution.
The main decay from the 3s 3p$^5$ 4s $^2$P$_{3/2}$, the strongest line, 
was tentatively identified in \cite{delzanna_etal:12_fe_10} with the 
96.0~\AA\ transition.

\def\baselinestretch{1.}

\begin{table*}[!htbp]
\caption{The relative intensities of a few soft X-ray lines in \ion{Fe}{ix}.}
\begin{center}
%\scriptsize
\footnotesize
\begin{tabular}[c]{@{}llcrccrccccl@{}}
\hline\hline\noalign{\smallskip}  
 $i$--$j$ & Levels   &  $Int$ &  $Int$ & $gf$ &  A$_{ji}$(s$^{-1}$) &   $\lambda_{\rm exp}$(\AA) & $\lambda_{\rm th}$(\AA)   & New \\   
  &  & 1.0$\times$10$^{8}$ & 1.0$\times$10$^{12}$ &  &    &  \\ 
\noalign{\smallskip}\hline\noalign{\smallskip}                                   %inserts single line

1--107 & 3s$^2$ 3p$^6$ $^1$S$_{0}$--3p$^5$ 4s $^1$P$_{1}$ & 1.0 & 0.76 & - & 4.1$\times$10$^{10}$ & 103.566 & 98.08 (-5.5) &   \\ 
\noalign{\smallskip}

5--302 & 3s$^2$ 3p$^5$ 3d $^3$F$_{4}$--3s$^2$ 3p$^5$ 5f $^3$G$_{5}$ & 0.14 & 0.12 & 2.07 & 1.6$\times$10$^{11}$ & 91.980 & 87.61 (-4.4) & ? 91.81  \\ 
5--366 & 3s$^2$ 3p$^5$ 3d $^3$F$_{4}$--3s$^2$ 3p$^5$ 6f $^3$G$_{5}$ & 7.7$\times$10$^{-2}$ & 6.7$\times$10$^{-2}$ & 1.64 & 1.6$\times$10$^{11}$ &  -  & 79.12 & ? 82.7  \\ 

13--276 & 3s$^2$ 3p$^5$ 3d $^1$P$_{1}$--3s$^2$ 3p$^5$ 5p $^1$S$_{0}$ & 5.6$\times$10$^{-2}$ & 3.2$\times$10$^{-2}$ & 6.5$\times$10$^{-3}$ & 3.2$\times$10$^{9}$ &  -  & 113.08 & ? 119.0  \\ 
10--271 & 3s$^2$ 3p$^5$ 3d $^3$D$_{1}$--3s$^2$ 3p$^5$ 5p $^3$P$_{0}$ & 5.4$\times$10$^{-2}$ & 3.1$\times$10$^{-2}$ & 4.9$\times$10$^{-2}$ & 3.1$\times$10$^{10}$ &  -  & 100.76 & ? 105.24  \\ 

13--326 & 3s$^2$ 3p$^5$ 3d $^1$P$_{1}$--3s$^2$ 3p$^5$ 5f $^1$D$_{2}$ & 4.9$\times$10$^{-2}$ & 3.3$\times$10$^{-2}$ & 0.73 & 9.5$\times$10$^{10}$ &  -  & 99.51 & ? 104.93  \\ 
13--379 & 3s$^2$ 3p$^5$ 3d $^1$P$_{1}$--3s$^2$ 3p$^5$ 6f $^1$D$_{2}$ & 4.6$\times$10$^{-2}$ & 3.0$\times$10$^{-2}$ & 0.50 & 8.4$\times$10$^{10}$ &  -  & 88.30 &  ? 92.75  \\ 
10--276 & 3s$^2$ 3p$^5$ 3d $^3$D$_{1}$--3s$^2$ 3p$^5$ 5p $^1$S$_{0}$ & 4.6$\times$10$^{-2}$ & 2.6$\times$10$^{-2}$ & 4.0$\times$10$^{-3}$ & 2.7$\times$10$^{9}$ &  -  & 99.18 &  ? 103.70  \\ 
10--316 & 3s$^2$ 3p$^5$ 3d $^3$D$_{1}$--3s$^2$ 3p$^5$ 5f $^3$F$_{2}$ & 3.5$\times$10$^{-2}$ & 2.4$\times$10$^{-2}$ & 0.44 & 7.2$\times$10$^{10}$ &  -  & 89.68 &  ? 94.15 \\ 

\noalign{\smallskip}\hline\noalign{\smallskip}                                   %inserts single line
\end{tabular}
\normalsize
\tablefoot{The relative line intensities  (photons)  $Int=N_{j} A_{ji}/N_{\rm e}$ 
were calculated at log N$_{\rm e}$ [cm$^{-3}$]=8,12 and log $T$$_{\rm e}$ [K]= 5.85
}
\end{center}
\label{tab:fe_9}
\end{table*}
% Table~\ref{tab:fe_9}

%\def\baselinestretch{1.5}

The Be72 intensity for this line is a bit low, however there are no
other strong lines in the vicinity. The 96.0+96.1~\AA\ blend, observed
by M72, has  a calibrated intensity in excellent agreement 
(within 30\%) with the 
quiet Sun PEVE value for the EUV 174.5~\AA\ line, as shown in  Fig.~\ref{fig:fe_10}.
The 94.0~\AA\ line also has an excellent agreement, if one assigns 
30\% of the M72 intensity to the 93.838~\AA\ (unidentified) line, as observed by 
Be72. 
This comparison confirms the accuracy (at least to about 30\%) of the new atomic
calculations.

Fawcett identified a few decays from the 3s$^2$ 3p$^4$ 4p. 
Based on this, we tentatively identify the 3--302 transition with a weak line
in the M72 spectrum at 109.52~\AA. The 27--302 transition would be at 151.42~\AA, 
where there is a weak line in the MH73 spectrum.
An alternative for the  3--302 transition would be the 108.53~\AA\ line.

\begin{figure}[!htbp]
\centerline{\epsfig{file=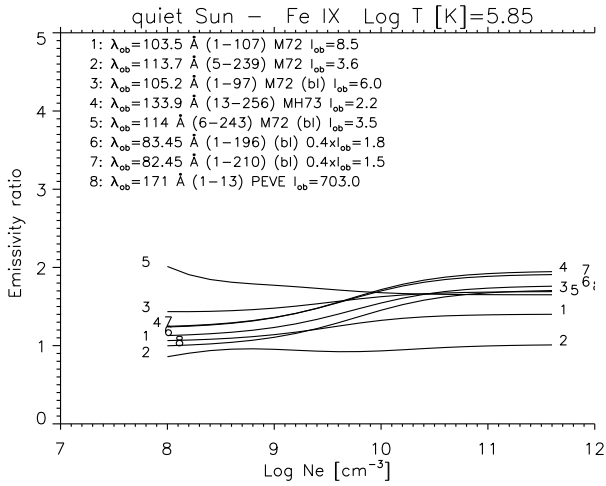, width=6.5cm,angle=90 }}
\caption{Emissivity ratio  curves  relative to  the 
main \ion{Fe}{ix} lines for the quiet Sun.}
 \label{fig:fe_9}
\end{figure}
% Fig.~\ref{fig:fe_9}

\subsection{\ion{Fe}{ix}}

The atomic data for  \ion{Fe}{ix} have been discussed in \cite{odwyer_etal:11_fe_9},
where the detailed list of the strongest lines 
can be found. The excitation rates for the 
3s$^2$ 3p$^5$ 4s and 3s$^2$ 3p$^5$ 4p levels are from \cite{storey_etal:02},
the rest from \cite{odwyer_etal:11_fe_9}.
The previous identifications are due to \cite{kruger_etal:37} (the two strong 
decays from the 4s levels),  \cite{alexander_etal:65}
(four decays from the 4d and 5s levels), and \cite{wagner_house:71} 
(12 transitions within the 3p$^5$ 3d -- 3p$^5$ 4f array).

 Fig.~\ref{fig:fe_9} shows the emissivity ratio  curves  relative to  the  M72 observations.
There is excellent (within $\pm$20\%) agreement among
all the brightest soft X-ray lines, and the EUV 171~\AA\ PEVE quiet Sun irradiance.
The weaker lines are blended at the M72 resolution, as noted in  Fig.~\ref{fig:fe_9}.

A few 3d--5f transitions were tentatively identified in 
\cite{odwyer_etal:11_fe_9}.
The strongest one is the 
 3s$^2$ 3p$^5$ 3d $^3$F$_{4}$--3s$^2$ 3p$^5$ 5f $^3$G$_{5}$ line,
which was predicted to be at 92.~\AA. 
The intensity of this line does not agree well with the calibrated M72 spectrum,
however, so it is suggested that this line blends the stronger solar line at 
91.81~\AA. A few new tentative identifications are proposed here
in Table~\ref{tab:fe_9}.

\subsection{An overall comparison}

The new atomic models provide intensities for a few millions of 
spectral lines in the soft X-rays. Given that this spectral 
region is  inherently 
over-crowded, we have also computed spectra to be compared to the 
observed ones, to see how much blending occurs from this forest of lines..

\begin{figure*}[!htbp]
\centerline{\epsfig{file=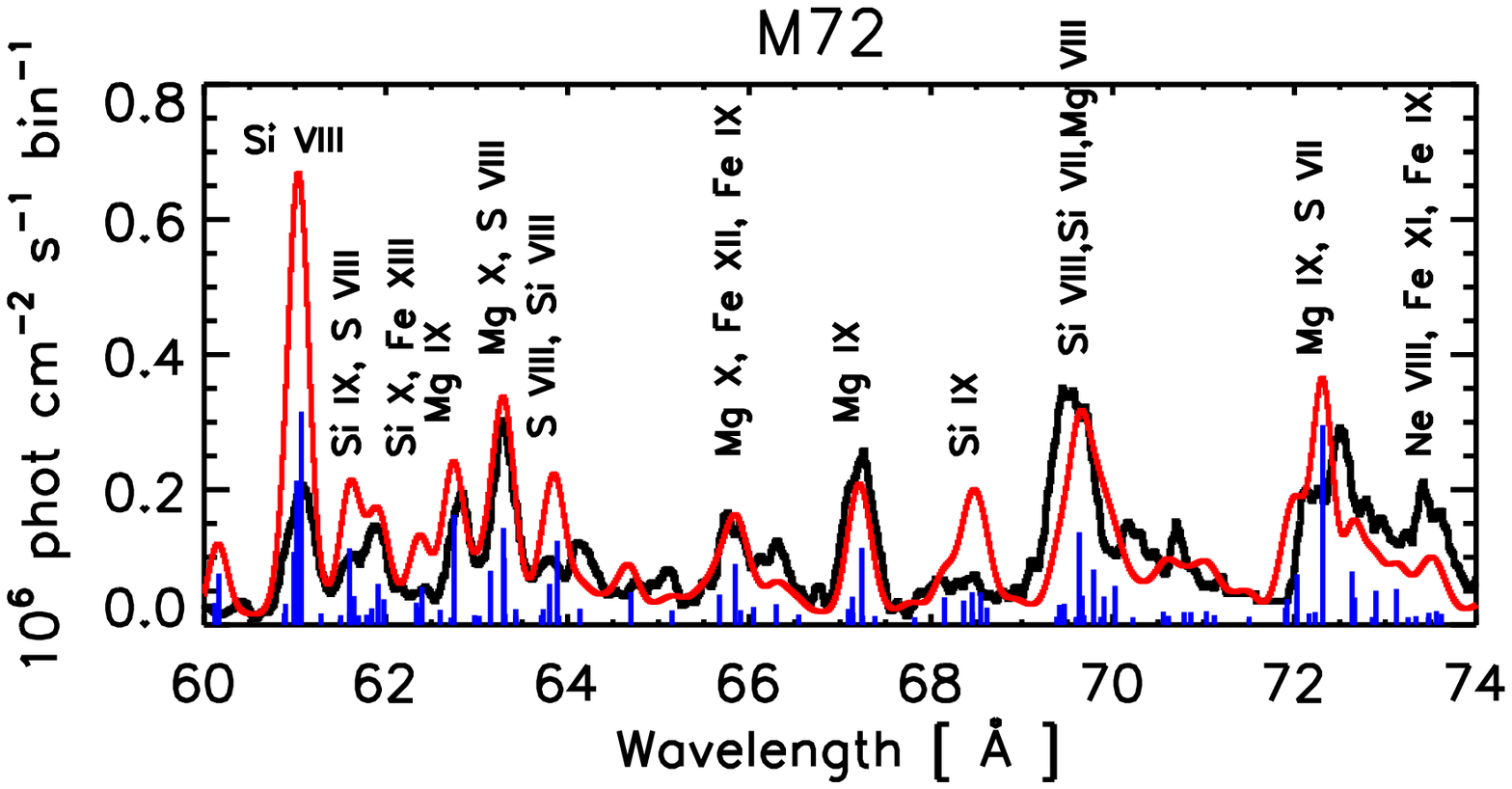, width=9.5cm,angle=0 }
\epsfig{file=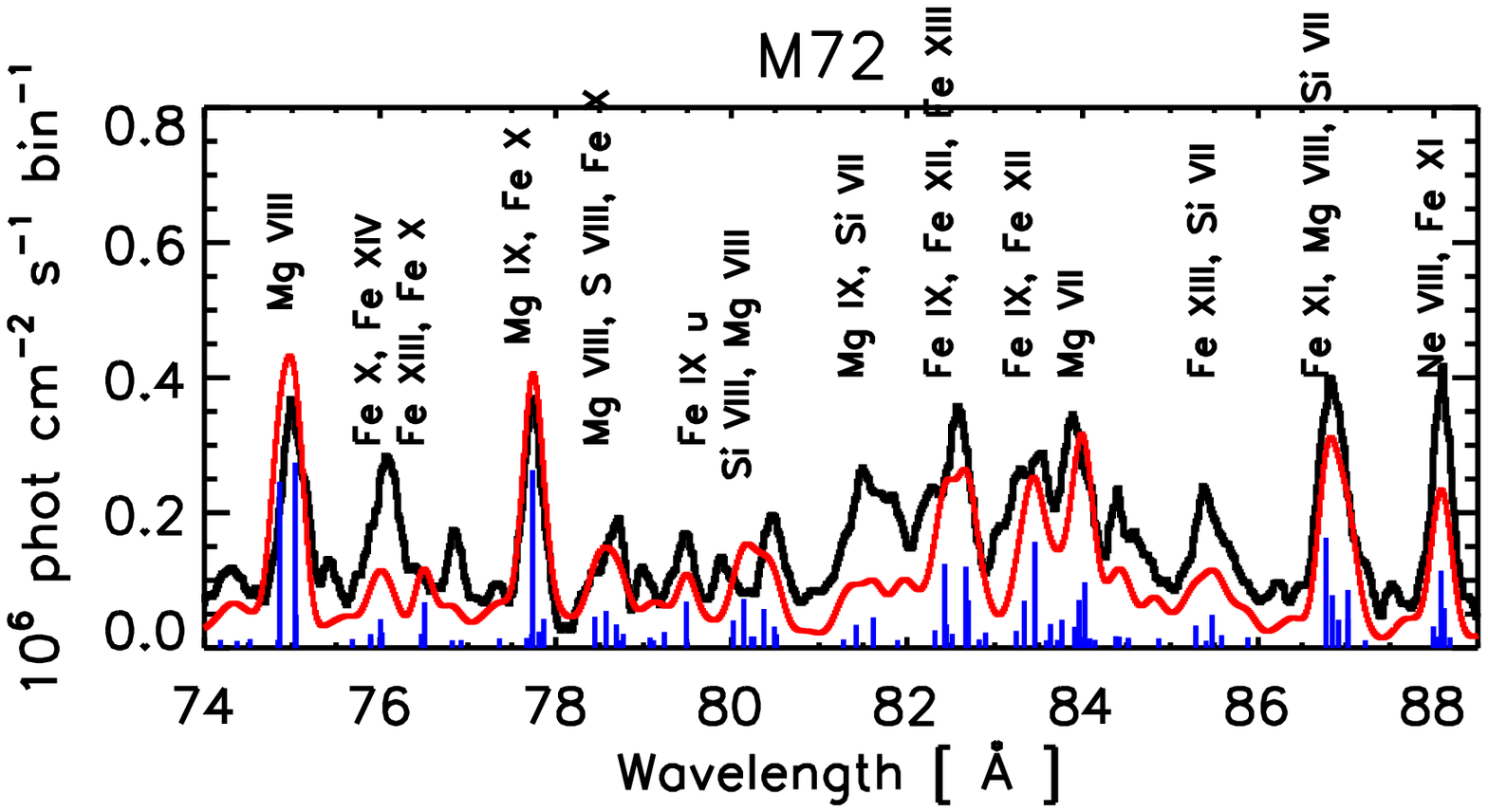, width=9.5cm,angle=0 }}
\centerline{\epsfig{file=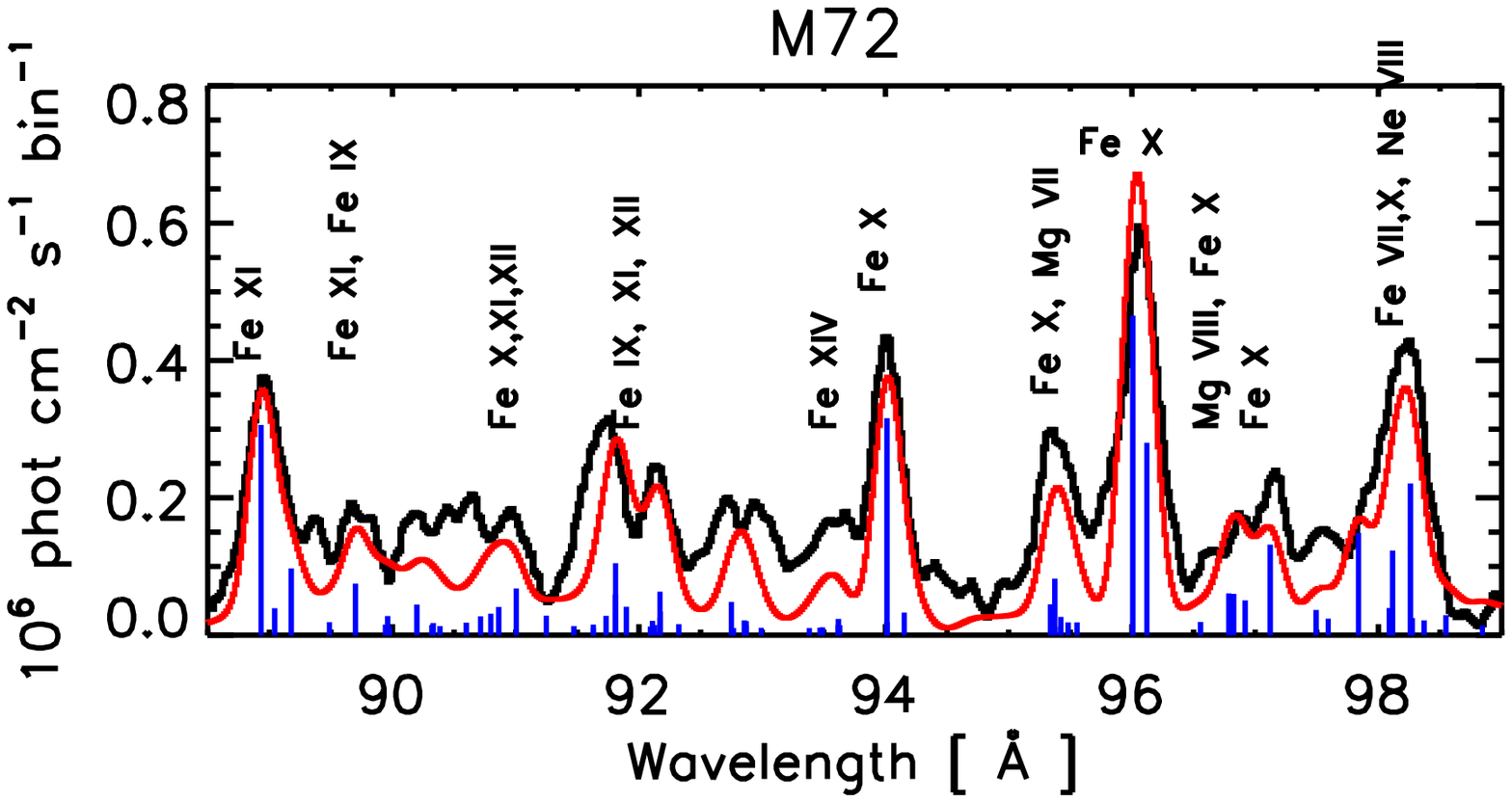, width=9.5cm,angle=0 }
\epsfig{file=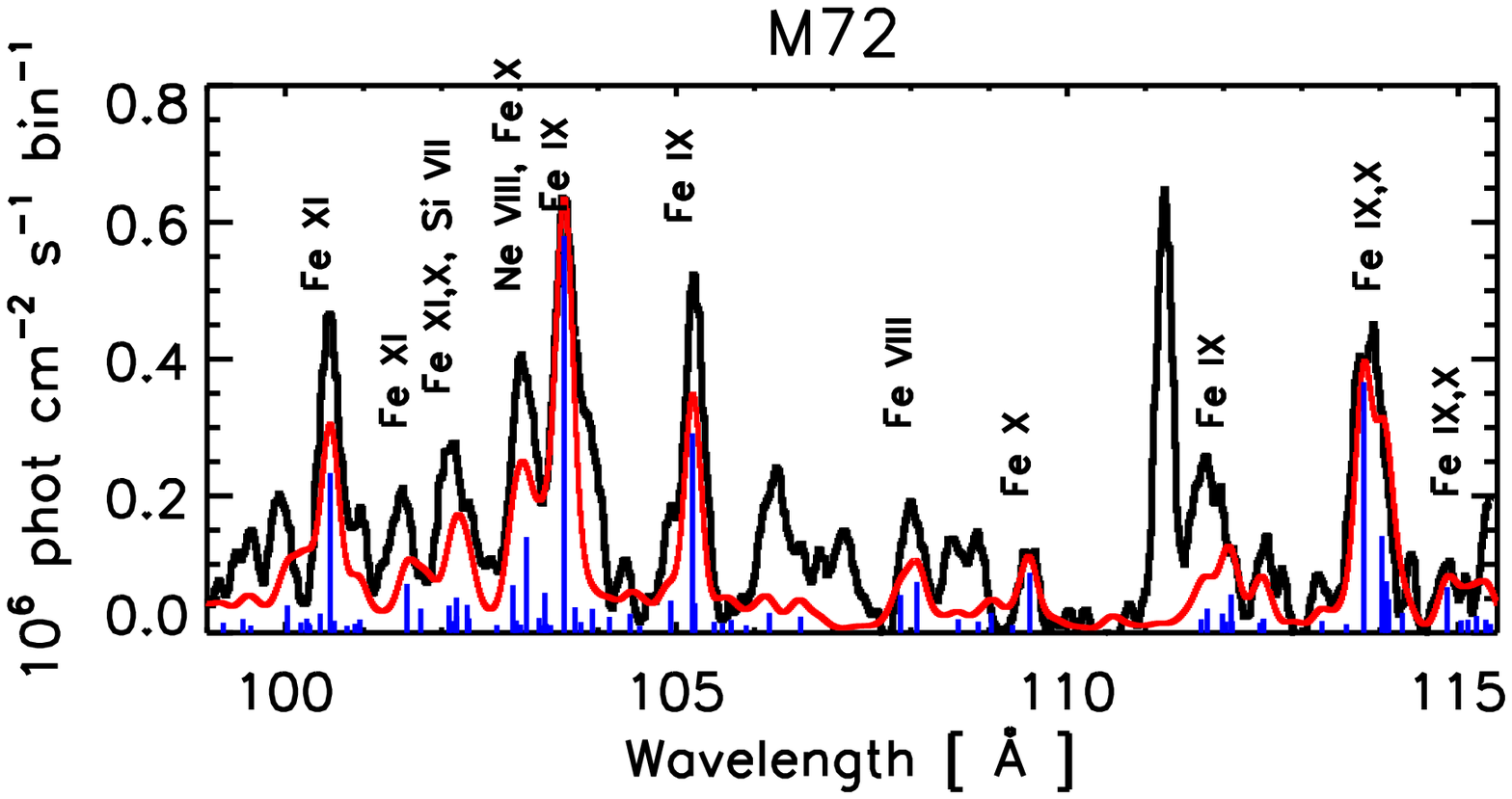, width=9.5cm,angle=0 }
}
\caption{Quiet Sun spectrum from M72, recalibrated (thick black),
with overplotted a theoretical spectrum (thin red).
The locations and intensities of the main lines are shown (blue
vertical lines). 
}
 \label{fig:m72_v7p}
\end{figure*}

We have taken the quiet Sun M72 spectrum and we have calibrated it
in wavelength, using the best known isolated and strong lines.
We have then flux-calibrated it, by comparing it with the M72 
published intensities and the PEVE ones. The resulting spectrum is shown in 
 Fig.~\ref{fig:m72_v7p} (black).

We have then adopted the set of `best' energies as calculated
for each of the iron ions. They were obtained by linear 
interpolation of the few known energies with respect to the 
target energies. We have added the identifications of the 
strongest lines provided here.
We have merged these datasets with those for all the other
ions not discussed here, using CHIANTI v.7 \citep{landi_etal:11_chianti_v7},
and computed line emissivities for quiet Sun conditions, at a constant
electron density of 1.0$\times$10$^{8}$  [cm$^{-3}$].

In order to obtain quiet Sun irradiances, we have folded the 
line emissivities with a quiet Sun differential emission measure 
(DEM) obtained from SOHO/CDS 
radiances not far from the limb by \cite{andretta_etal:03}, assuming photospheric abundances.
For the forward modelling, we have adopted the new CHIANTI 
ion fractions, and a recent  set of photospheric abundances
by \cite{asplund_etal:09}. 
We have then roughly converted radiances into irradiances by neglecting 
limb-brightening and  off-limb contributions.
A proper treatment would just scale the absolute values of the 
irradiances. 
The irradiances have then been folded with Gaussian line profiles 
to match the M72 observed spectra,
and put onto a wavelength grid with a bin size similar to the 
M72 one. The resulting spectra are also shown in  Fig.~\ref{fig:m72_v7p}
(red). The agreement is remarkable.
The same figure also shows (in blue) the location and intensities of the main 
lines contributing to the calculated spectra. This clearly shows,
as we knew, that the majority of the lines at the M72 resolution
are blends of many transitions. The most significant ones are labelled.

% Fig.~\ref{fig:m72_v7p}

\begin{figure}[!htbp]
\centerline{\epsfig{file=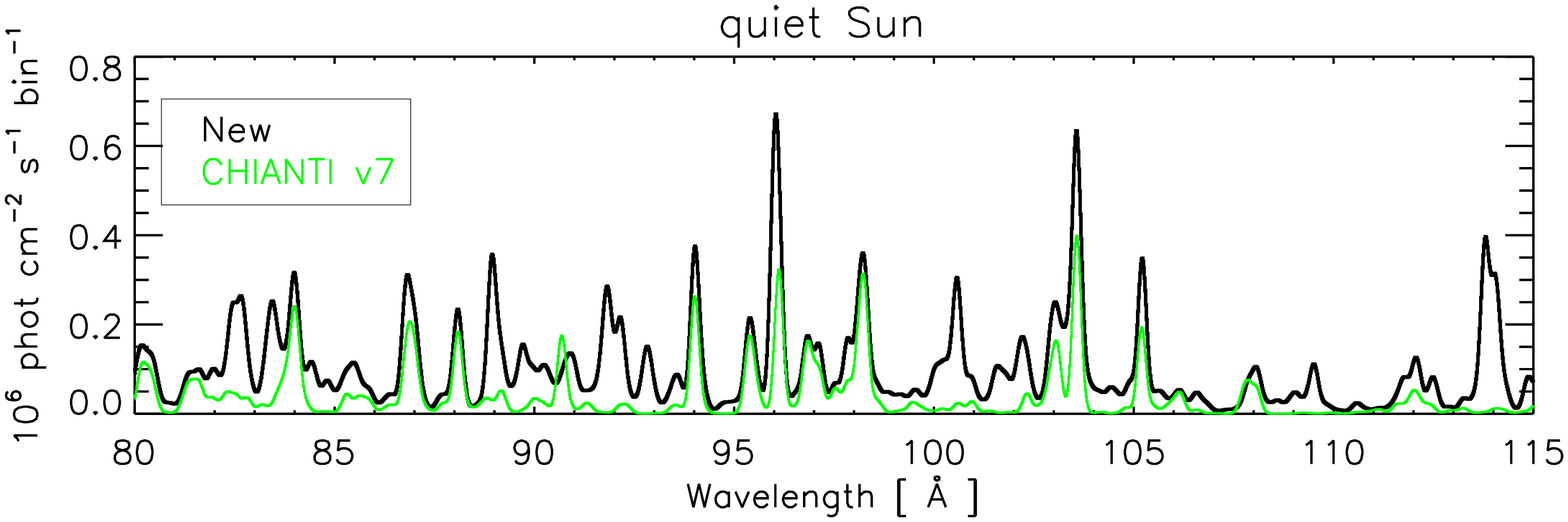, width=9.5cm,angle=0}}
\centerline{\epsfig{file=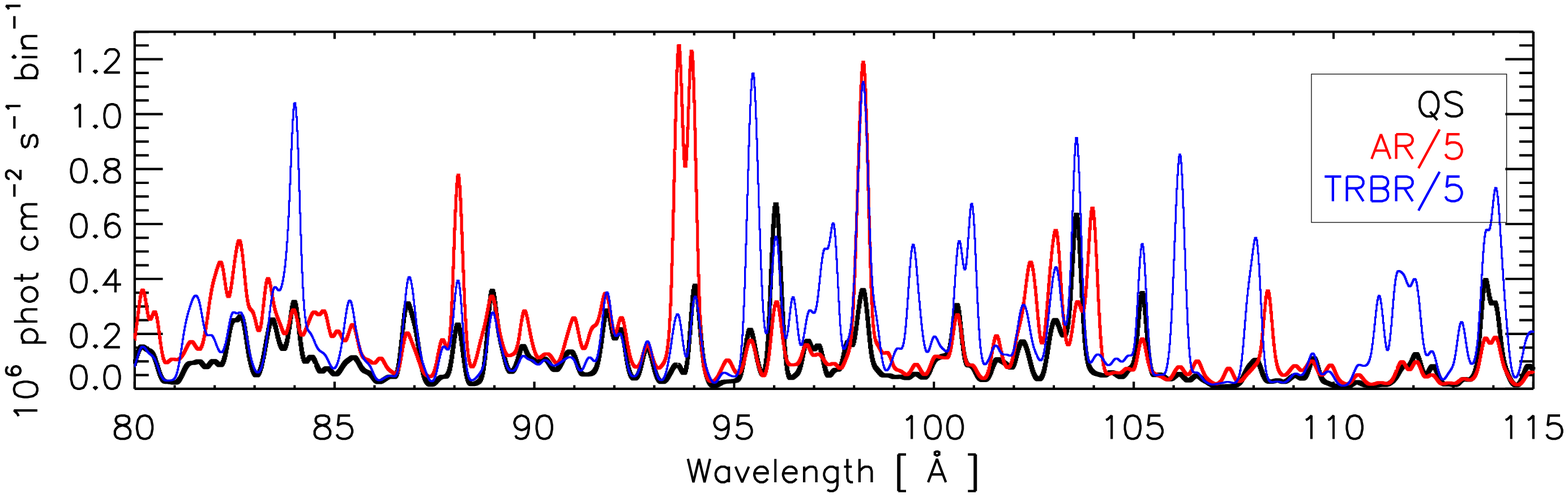, width=9.5cm,angle=0}}
\caption{Top: quiet Sun theoretical spectra, from the present 
atomic data and identifications, compared to CHIANTI v.7.
Bottom: theoretical spectra for the quiet Sun (QS, thickest black),
an active region (AR, reduced by a factor of 5, thick red), and the 
base of a loop (TRBR,  reduced by a factor of 5, blue thin).
}
 \label{fig:spectra}
\end{figure}
% Fig.~\ref{fig:spectra}

The 80--106~\AA\ spectral region is dominated by iron ions, indeed
 Fig.~\ref{fig:spectra} (left) shows the improvement with the new atomic data,
compared to CHIANTI v.7.
Fig.~\ref{fig:m72_v7p} shows that 
 a significant fraction of flux is still  missing, being probably 
due to a range of other ions which are emitting at these wavelengths.
The 60--80~\AA\ spectral region in the quiet Sun is dominated by
a range of non-iron ions instead.  Relatively good agreement in terms of 
wavelengths and intensities is present, with two notable  exceptions, 
several lines from \ion{Si}{viii} and \ion{Si}{ix} around 61 and 68~\AA.

Different source regions of the solar corona are going to produce very 
different spectra. 
This is an additional complication for the benchmark and for the 
analysis of solar spectra.
To show how different they are, an average active region spectrum has been obtained
from a DEM based on the SERTS-89 observation \citep{thomas_neupert:94,delzanna_thesis99}.
This is shown in  Fig.~\ref{fig:spectra} (right, AR).
In order to assess the contributions due to the cooler transition-region ($T \le$ 1~MK)
lines, a spectrum of the 
base of an active region loop (region B in \citealt{delzanna_etal:11_aia}) has been calculated,
and is also shown in  Fig.~\ref{fig:spectra} (right, TRBR).

\subsection{The SDO AIA 94~\AA\ band}

\begin{figure}[!htbp]
\centerline{\epsfig{file=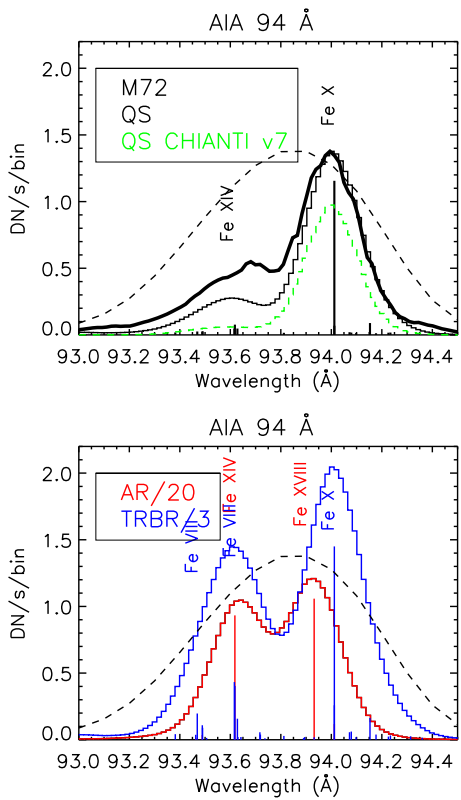, width=7.cm,angle=0 }}
\caption{SDO AIA 94~\AA\ simulated count rates, obtained for the quiet Sun (QS),
an active region core (AR, reduced by a factor of 20) and a loop base 
(TRBR,  reduced by a factor of 3). The M72 quiet Sun spectrum 
(normalised) is also shown, as well as the quiet Sun spectrum obtained from CHIANTI v7.
The AIA effective area (normalised) is shown as a dash line. }
 \label{fig:aia_94}
\end{figure}

The  SDO AIA 94~\AA\ band has been the subject of various studies, to try an 
resolve the discrepancies in terms of atomic data
\citep{delzanna_etal:11_aia,odwyer_etal:11_fe_9,foster_testa:11, testa_etal:12}. 
In order to see how the new atomic data and identifications affect
the SDO AIA 94~\AA\ band, the predicted quiet Sun spectrum 
has been folded with the AIA effective area, to provide estimated 
count rates per AIA pixel.
They are shown in Fig.~\ref{fig:aia_94} (black thin spectrum).
For comparison, a normalised M72 spectrum is overplotted (thick black spectrum),
as well as what is calculated with the previous CHIANTI v7 (green dashed), which 
had, for \ion{Fe}{x}, incorrect atomic data. 
The plot shows that, for the quiet Sun, some missing flux is still 
present, in the blue wing of the dominant contribution from \ion{Fe}{x}. 
This was expected. 
Be72 reports four strong lines at 
93.618, 93.838, 93.933, and 94.016~\AA. 
The latter line is only 4 m~\AA\ long-ward of Edlen's measurement of 
94.012 for the \ion{Fe}{x} line.
The 93.933~\AA\ has the same wavelength of the 
strong \ion{Fe}{xviii} line at 93.932~\AA\  \citep{delzanna:06_fe_18},
and is likely that indeed this ion provides the observed counts.
The 93.618\AA\  has been identified here as \ion{Fe}{xiv}
(blended with \ion{Fe}{viii}), but the 
 93.838 still remains unidentified.
There is a line also present in Fawcett C53 plate at the same wavelength,
which could be a coincidence or the same  transition due to an iron ion.
Further blending of weaker lines is possible.

The atomic data for a range of ions which produce lines observed in the 
laboratory  or predicted to fall around these wavelengths
have been assessed, but no significant missing flux in solar 
conditions has yet been found. 
For example, strong lines from \ion{Mg}{vii} and 
 \ion{Mg}{viii} have been observed at 94.043~\AA. 
A few transitions from \ion{Al}{v} are also present, however the
APAP atomic data \citep{witthoeft_etal:07_f-like} indicate that they would be weak.
There is an  \ion{O}{vi} 2s--9p observed at 93.84~\AA,
however even the presence of the 2s--8p and 2s--7p  transitions
is dubious. 
 New atomic calculations for some ions are in progress to clarify this.

The newly identified \ion{Fe}{xiv} at 93.61~\AA\ does provide a
significant contribution to the 94~\AA\ band even for the quiet Sun. 
This becomes even more significant in active regions, as  
shown  in Fig.~\ref{fig:aia_94} (AR, reduced by a factor of 20 and 
obtained from the SERTS-89 observation).
The dominant count rates in the band are originating from \ion{Fe}{xiv}
and the \ion{Fe}{xviii} line at 93.932~\AA.

In order to assess the contributions due to the cooler
\ion{Fe}{viii} and \ion{Fe}{ix} lines, the spectrum of the 
base of an active region loop (see above)
is also shown in  Fig.~\ref{fig:aia_94} (blue, reduced by a factor of 3). 
Indeed in this particular case, the \ion{Fe}{viii} transitions \citep{odwyer_etal:11_fe_9}
produce a significant contribution to this band.

\cite{delzanna_etal:11_aia}
presented a detailed comparison of SDO AIA and Hinode/EIS spectra, 
showing that indeed there are for the 94~\AA\ band, aside from the 1~MK \ion{Fe}{x}
contribution, at least two additional  components.
One is a hot component, which we identify with \ion{Fe}{xiv}, and one is 
a cooler component, which we identify with  \ion{Fe}{viii}.

% Fig.~\ref{fig:aia_94}

\section{Conclusions}

This paper is the first benchmark for the soft X-ray lines.
It is a summary of almost two years of work on the calculations
and identifications of the  soft X-ray lines due to $n=4 \to n=3$ transitions 
of the main iron ions.
Large-scale  R-matrix and distorted wave  scattering calculations turned out to be 
both needed, to account
for resonance enhancements in the excitation rates for the $n=4$ levels, and for cascading
from higher levels.

The identification work proved very difficult, due to the lack of
high-resolution well-calibrated spectra, the fact that 
the soft X-rays are notoriously packed with a large number
of transitions from a range of ions, and that laboratory spectra and solar
spectra are very much different.

The strongest iron transitions are all finally identified here.
Very good agreement between the soft-Xray ($n=4 \to n=3$) and EUV
($n=3 \to n=3$) irradiances of the strongest lines
 is found for the first time, confirming the reliability
of the new calculations. 

In several cases, various discrepancies in the previous identifications have been found,
and many tentative (new or revised) identifications have been proposed.
 Better experimental data and more atomic 
calculations on a range of other ions will be needed to confirm them.
Some  calculations for other ions that produce strong 
lines in the soft X-rays  are already in progress.

With regard to the SDO AIA 94~\AA\ band, good progress
has been made, with a new important identification of a
strong  \ion{Fe}{xiv} line at 93.61~\AA, and the new calculations
for  \ion{Fe}{x}, \ion{Fe}{ix} and \ion{Fe}{viii}.
 At least one residual transition still need to be identified though.

The new  large amount of APAP atomic data  will be made available
through the CHIANTI database, however this will require a new format 
and a new way to handle them. Work is in progress in this direction.

\begin{acknowledgements}
I acknowledge  STFC (UK) support  via the Advanced Fellowship programme
and the APAP network.
 B.C. Fawcett  is thanked for his contribution in rescuing some of his
original plates, and for the continuous encouragement over the years.
P.J. Storey and H.E.Mason are also thanked for useful discussions.

\end{acknowledgements}

\bibliographystyle{aa}

%\bibliography{../bib}
\bibliography{H19923}

\begin{thebibliography}{63}
\expandafter\ifx\csname natexlab\endcsname\relax\def\natexlab#1{#1}\fi

\bibitem[{{Acton} {et~al.}(1985){Acton}, {Bruner}, {Brown}, {Fawcett},
  {Schweizer}, \& {Speer}}]{acton_etal:85}
{Acton}, L.~W., {Bruner}, M.~E., {Brown}, W.~A., {et~al.} 1985, \apj, 291, 865

\bibitem[{{Aggarwal} \& {Keenan}(2006)}]{aggarwal_keenan:06}
{Aggarwal}, K.~M. \& {Keenan}, F.~P. 2006, \aap, 450, 1249

\bibitem[{{Aggarwal} {et~al.}(2003){Aggarwal}, {Keenan}, \&
  {Msezane}}]{aggarwal_etal:03}
{Aggarwal}, K.~M., {Keenan}, F.~P., \& {Msezane}, A.~Z. 2003, \aap, 410, 349

\bibitem[{{Alexander} {et~al.}(1965){Alexander}, {Feldman}, \&
  {Fraenkel}}]{alexander_etal:65}
{Alexander}, E., {Feldman}, U., \& {Fraenkel}, B.~S. 1965, Journal of the
  Optical Society of America (1917-1983), 55, 650

\bibitem[{{Andretta} {et~al.}(2003){Andretta}, {Del Zanna}, \&
  {Jordan}}]{andretta_etal:03}
{Andretta}, V., {Del Zanna}, G., \& {Jordan}, S.~D. 2003, \aap, 400, 737

\bibitem[{{Asplund} {et~al.}(2009){Asplund}, {Grevesse}, {Sauval}, \&
  {Scott}}]{asplund_etal:09}
{Asplund}, M., {Grevesse}, N., {Sauval}, A.~J., \& {Scott}, P. 2009, \araa, 47,
  481

\bibitem[{{Behring} {et~al.}(1972){Behring}, {Cohen}, \&
  {Feldman}}]{behring_etal:72}
{Behring}, W.~E., {Cohen}, L., \& {Feldman}, U. 1972, \apj, 175, 493

\bibitem[{{Berrington} {et~al.}(2005){Berrington}, {Ballance}, {Griffin}, \&
  {Badnell}}]{berrington_etal:05_fe_15}
{Berrington}, K.~A., {Ballance}, C.~P., {Griffin}, D.~C., \& {Badnell}, N.~R.
  2005, Journal of Physics B Atomic Molecular Physics, 38, 1667

\bibitem[{{Bhatia} {et~al.}(1997){Bhatia}, {Mason}, \&
  {Blancard}}]{bhatia_etal:97}
{Bhatia}, A.~K., {Mason}, H.~E., \& {Blancard}, C. 1997, Atomic Data and
  Nuclear Data Tables, 66, 83

\bibitem[{{Chamberlin} {et~al.}(2009){Chamberlin}, {Woods}, {Crotser},
  {Eparvier}, {Hock}, \& {Woodraska}}]{chamberlin_etal:09}
{Chamberlin}, P.~C., {Woods}, T.~N., {Crotser}, D.~A., {et~al.} 2009, \grl, 36,
  5102

\bibitem[{{Cornille} {et~al.}(1997){Cornille}, {Dubau}, {Mason}, {Blancard}, \&
  {Brown}}]{cornille_etal:97}
{Cornille}, M., {Dubau}, J., {Mason}, H.~E., {Blancard}, C., \& {Brown}, W.~A.
  1997, \aap, 320, 333

\bibitem[{{Cowan} \& {Widing}(1973)}]{cowan_widing:73}
{Cowan}, R.~D. \& {Widing}, K.~G. 1973, \apj, 180, 285

\bibitem[{{Del Zanna}(1999)}]{delzanna_thesis99}
{Del Zanna}, G. 1999, PhD thesis, Univ.\ of Central Lancashire, UK

\bibitem[{{Del Zanna}(2006)}]{delzanna:06_fe_18}
{Del Zanna}, G. 2006, \aap, 459, 307

\bibitem[{{Del Zanna}(2010)}]{delzanna:10_fe_11}
{Del Zanna}, G. 2010, \aap, 514, A41+

\bibitem[{{Del Zanna}(2011)}]{delzanna:11_fe_13}
{Del Zanna}, G. 2011, \aap, 533

\bibitem[{{Del Zanna} \& {Andretta}(2011)}]{delzanna_andretta:11}
{Del Zanna}, G. \& {Andretta}, V. 2011, \aap, 528, A139

\bibitem[{{Del Zanna} {et~al.}(2010){Del Zanna}, {Andretta}, {Chamberlin},
  {Woods}, \& {Thompson}}]{delzanna_etal:10_cdscal}
{Del Zanna}, G., {Andretta}, V., {Chamberlin}, P.~C., {Woods}, T.~N., \&
  {Thompson}, W.~T. 2010, \aap, 518, A49

\bibitem[{{Del Zanna} {et~al.}(2004){Del Zanna}, {Berrington}, \&
  {Mason}}]{delzanna_etal:04_fe_10}
{Del Zanna}, G., {Berrington}, K.~A., \& {Mason}, H.~E. 2004, \aap, 422, 731

\bibitem[{{Del Zanna} \& {Mason}(2005)}]{delzanna_mason:05_fe_12}
{Del Zanna}, G. \& {Mason}, H.~E. 2005, \aap, 433, 731

\bibitem[{{Del Zanna} {et~al.}(2011){Del Zanna}, {O'Dwyer}, \&
  {Mason}}]{delzanna_etal:11_aia}
{Del Zanna}, G., {O'Dwyer}, B., \& {Mason}, H.~E. 2011, \aap, 535, A46

\bibitem[{{Del Zanna} {et~al.}(2008){Del Zanna}, {Rozum}, \&
  {Badnell}}]{delzanna_etal:08_mg_9}
{Del Zanna}, G., {Rozum}, I., \& {Badnell}, N. 2008, \aap, 487, 1203

\bibitem[{{Del Zanna} \&
  {Storey}(2012{\natexlab{a}})}]{delzanna_storey:12_fe_11}
{Del Zanna}, G. \& {Storey}, P.~J. 2012{\natexlab{a}}, \aap, submitted

\bibitem[{{Del Zanna} \&
  {Storey}(2012{\natexlab{b}})}]{delzanna_storey:12_fe_13}
{Del Zanna}, G. \& {Storey}, P.~J. 2012{\natexlab{b}}, \aap, 543, A144

\bibitem[{{Del Zanna} {et~al.}(2012{\natexlab{a}}){Del Zanna}, {Storey},
  {Badnell}, \& {Mason}}]{delzanna_etal:12_fe_12}
{Del Zanna}, G., {Storey}, P.~J., {Badnell}, N.~R., \& {Mason}, H.~E.
  2012{\natexlab{a}}, \aap, 543, A139

\bibitem[{{Del Zanna} {et~al.}(2012{\natexlab{b}}){Del Zanna}, {Storey},
  {Badnell}, \& {Mason}}]{delzanna_etal:12_fe_10}
{Del Zanna}, G., {Storey}, P.~J., {Badnell}, N.~R., \& {Mason}, H.~E.
  2012{\natexlab{b}}, \aap, 541, A90

\bibitem[{{Dere} {et~al.}(1997){Dere}, {Landi}, {Mason}, {Monsignori Fossi}, \&
  {Young}}]{dere_etal:97}
{Dere}, K.~P., {Landi}, E., {Mason}, H.~E., {Monsignori Fossi}, B.~C., \&
  {Young}, P.~R. 1997, \aaps, 125, 149

\bibitem[{{Dere} {et~al.}(2009){Dere}, {Landi}, {Young}, {Del Zanna}, {Mason},
  \& {Landini}}]{dere_etal:09_chianti_v6}
{Dere}, K.~P., {Landi}, E., {Young}, P.~R., {et~al.} 2009, \aap, 498, 915

\bibitem[{{Edl{\' e}n}(1937{\natexlab{a}})}]{edlen:37b}
{Edl{\' e}n}, B. 1937{\natexlab{a}}, Zeitschrift fur Astrophysics, 104, 407

\bibitem[{{Edl{\' e}n}(1937{\natexlab{b}})}]{edlen:37_s-like}
{Edl{\' e}n}, B. 1937{\natexlab{b}}, Zeitschrift fur Astrophysics, 104, 188

\bibitem[{{Edl{\' e}n}(1937{\natexlab{c}})}]{edlen:37_cl-like}
{Edl{\' e}n}, B. 1937{\natexlab{c}}, Zeitschrift f\"ur Astrophysics, 104, 407

\bibitem[{{Edl{\'e}n}(1936{\natexlab{a}})}]{edlen:36_mg-like}
{Edl{\'e}n}, B. 1936{\natexlab{a}}, Zeitschrift fur Physik, 103, 536

\bibitem[{{Edl{\'e}n}(1936{\natexlab{b}})}]{edlen:36_na-like}
{Edl{\'e}n}, B. 1936{\natexlab{b}}, Zeitschrift fur Physik, 100, 621

\bibitem[{{Fawcett} {et~al.}(1972){Fawcett}, {Kononov}, {Hayes}, \&
  {Cowan}}]{fawcett_etal:72}
{Fawcett}, B.~C., {Kononov}, E.~Y., {Hayes}, R.~W., \& {Cowan}, R.~D. 1972,
  Journal of Physics B Atomic Molecular Physics, 5, 1255

\bibitem[{{Feldman} {et~al.}(1970){Feldman}, {Cohen}, \&
  {Behring}}]{feldman_etal:70_li-like}
{Feldman}, U., {Cohen}, L., \& {Behring}, W. 1970, Journal of the Optical
  Society of America (1917-1983), 60, 891

\bibitem[{{Foster} \& {Testa}(2011)}]{foster_testa:11}
{Foster}, A.~R. \& {Testa}, P. 2011, \apjl, 740, L52

\bibitem[{{Kastner} {et~al.}(1978){Kastner}, {Swartz}, {Bhatia}, \&
  {Lapides}}]{kastner_etal:78}
{Kastner}, S.~O., {Swartz}, M., {Bhatia}, A.~K., \& {Lapides}, J. 1978, J. Opt.
  Soc. Am., 68, 1558

\bibitem[{{Keenan} {et~al.}(2007){Keenan}, {Drake}, \&
  {Aggarwal}}]{keenan_etal:07}
{Keenan}, F.~P., {Drake}, J.~J., \& {Aggarwal}, K.~M. 2007, \mnras, 381, 1727

\bibitem[{{Keenan} {et~al.}(2006){Keenan}, {Drake}, {Chung}, {Brickhouse},
  {Aggarwal}, {Msezane}, {Ryans}, \& {Bloomfield}}]{keenan_etal:06}
{Keenan}, F.~P., {Drake}, J.~J., {Chung}, S., {et~al.} 2006, \apj, 645, 597

\bibitem[{{Kink} {et~al.}(1997){Kink}, {Tunklev}, \&
  {Litz{\'e}n}}]{kink_etal:97_fe_15}
{Kink}, I., {Tunklev}, M., \& {Litz{\'e}n}, U. 1997, Journal of the Optical
  Society of America B Optical Physics, 14, 722

\bibitem[{{Kruger} {et~al.}(1937){Kruger}, {Weissberg}, \&
  {Phillips}}]{kruger_etal:37}
{Kruger}, P.~G., {Weissberg}, S.~G., \& {Phillips}, L.~W. 1937, Physical
  Review, 51, 1090

\bibitem[{{Landi}(2011)}]{landi:2011_fe_15}
{Landi}, E. 2011, Atomic Data and Nuclear Data Tables, 97, 587

\bibitem[{{Landi} {et~al.}(2012){Landi}, {Del Zanna}, {Young}, {Dere}, \&
  {Mason}}]{landi_etal:11_chianti_v7}
{Landi}, E., {Del Zanna}, G., {Young}, P.~R., {Dere}, K.~P., \& {Mason}, H.~E.
  2012, \apj, 744, 99

\bibitem[{{Lemen} {et~al.}(2012){Lemen}, {Title}, {Akin}, {Boerner}, {Chou},
  {Drake}, {Duncan}, {Edwards}, {Friedlaender}, {Heyman}, {Hurlburt}, {Katz},
  {Kushner}, {Levay}, {Lindgren}, {Mathur}, {McFeaters}, {Mitchell}, {Rehse},
  {Schrijver}, {Springer}, {Stern}, {Tarbell}, {Wuelser}, {Wolfson}, {Yanari},
  {Bookbinder}, {Cheimets}, {Caldwell}, {Deluca}, {Gates}, {Golub}, {Park},
  {Podgorski}, {Bush}, {Scherrer}, {Gummin}, {Smith}, {Auker}, {Jerram},
  {Pool}, {Soufli}, {Windt}, {Beardsley}, {Clapp}, {Lang}, \&
  {Waltham}}]{lemen_etal:12}
{Lemen}, J.~R., {Title}, A.~M., {Akin}, D.~J., {et~al.} 2012, \solphys, 275, 17

\bibitem[{{Lepson} {et~al.}(2002){Lepson}, {Beiersdorfer}, {Brown}, {Liedahl},
  {Utter}, {Brickhouse}, {Dupree}, {Kaastra}, {Mewe}, \&
  {Kahn}}]{lepson_etal:02}
{Lepson}, J.~K., {Beiersdorfer}, P., {Brown}, G.~V., {et~al.} 2002, \apj, 578,
  648

\bibitem[{{Liang} {et~al.}(2010){Liang}, {Badnell}, {Crespo L{\'o}pez-Urrutia},
  {Baumann}, {Del Zanna}, {Storey}, {Tawara}, \&
  {Ullrich}}]{liang_etal:10_fe_14}
{Liang}, G.~Y., {Badnell}, N.~R., {Crespo L{\'o}pez-Urrutia}, J.~R., {et~al.}
  2010, \apjs, 190, 322

\bibitem[{{Liang} {et~al.}(2009){Liang}, {Whiteford}, \&
  {Badnell}}]{liang_etal:09_na-like}
{Liang}, G.~Y., {Whiteford}, A.~D., \& {Badnell}, N.~R. 2009, \aap, 500, 1263

\bibitem[{{Liang} \& {Zhao}(2010)}]{liang_zhao:10}
{Liang}, G.~Y. \& {Zhao}, G. 2010, \mnras, 405, 1987

\bibitem[{{Malinovsky} \& {Heroux}(1973)}]{malinovsky_heroux:73}
{Malinovsky}, L. \& {Heroux}, M. 1973, \apj, 181, 1009

\bibitem[{{Malinovsky} {et~al.}(1980){Malinovsky}, {Dubau}, \&
  {Sahal-Brechot}}]{malinovsky_etal:80}
{Malinovsky}, M., {Dubau}, J., \& {Sahal-Brechot}, S. 1980, \apj, 235, 665

\bibitem[{{Manson}(1972)}]{manson:72}
{Manson}, J.~E. 1972, \solphys, 27, 107

\bibitem[{{O'Dwyer} {et~al.}(2012){O'Dwyer}, {Del Zanna}, {Badnell}, {Mason},
  \& {Storey}}]{odwyer_etal:11_fe_9}
{O'Dwyer}, B., {Del Zanna}, G., {Badnell}, N.~R., {Mason}, H.~E., \& {Storey},
  P.~J. 2012, \aap, 537, A22

\bibitem[{{Sampson} {et~al.}(1990){Sampson}, {Zhang}, \&
  {Fontes}}]{sampson_et_al:90}
{Sampson}, D.~H., {Zhang}, H.~L., \& {Fontes}, C.~J. 1990, Atomic Data and
  Nuclear Data Tables, 44, 209

\bibitem[{{S{\"o}derqvist}(1944)}]{soderqvist:44_mg_9}
{S{\"o}derqvist}, J. 1944, J Ark. Mat. Astron. Fysik, 30

\bibitem[{{Storey} {et~al.}(2002){Storey}, {Zeippen}, \& {Le
  Dourneuf}}]{storey_etal:02}
{Storey}, P.~J., {Zeippen}, C.~J., \& {Le Dourneuf}, M. 2002, \aap, 394, 753

\bibitem[{{Testa} {et~al.}(2012){Testa}, {Drake}, \& {Landi}}]{testa_etal:12}
{Testa}, P., {Drake}, J.~J., \& {Landi}, E. 2012, \apj, 745, 111

\bibitem[{{Thomas} \& {Neupert}(1994)}]{thomas_neupert:94}
{Thomas}, R.~J. \& {Neupert}, W.~M. 1994, \apjs, 91, 461

\bibitem[{{Vilkas} \& {Ishikawa}(2004)}]{vilkas_ishikawa:04_fe_13n4}
{Vilkas}, M.~J. \& {Ishikawa}, Y. 2004, \pra, 69, 062503

\bibitem[{{Wagner} \& {House}(1971)}]{wagner_house:71}
{Wagner}, W.~J. \& {House}, L.~L. 1971, \apj, 166, 683

\bibitem[{{Witthoeft} {et~al.}(2007){Witthoeft}, {Whiteford}, \&
  {Badnell}}]{witthoeft_etal:07_f-like}
{Witthoeft}, M.~C., {Whiteford}, A.~D., \& {Badnell}, N.~R. 2007, Journal of
  Physics B Atomic Molecular Physics, 40, 2969

\bibitem[{{Woods} {et~al.}(2009){Woods}, {Chamberlin}, {Harder}, {Hock},
  {Snow}, {Eparvier}, {Fontenla}, {McClintock}, \& {Richard}}]{woods_etal:09}
{Woods}, T.~N., {Chamberlin}, P.~C., {Harder}, J.~W., {et~al.} 2009, \grl, 36,
  1101

\bibitem[{{Woods} {et~al.}(2012){Woods}, {Eparvier}, {Hock}, {Jones},
  {Woodraska}, {Judge}, {Didkovsky}, {Lean}, {Mariska}, {Warren}, {McMullin},
  {Chamberlin}, {Berthiaume}, {Bailey}, {Fuller-Rowell}, {Sojka}, {Tobiska}, \&
  {Viereck}}]{woods_etal:12}
{Woods}, T.~N., {Eparvier}, F.~G., {Hock}, R., {et~al.} 2012, \solphys, 275,
  115

\bibitem[{{Zhang} {et~al.}(1990){Zhang}, {Sampson}, \&
  {Fontes}}]{zhang_etal:90}
{Zhang}, H.~L., {Sampson}, D.~H., \& {Fontes}, C.~J. 1990, Atomic Data and
  Nuclear Data Tables, 44, 31

\end{thebibliography}

%\appendix

\end{document}